\documentclass[12pt]{article}
\usepackage{amsmath,amssymb,amsfonts,color,graphicx,cite,color,soul,ulem}
\newcount\timecount
\newcount\hours \newcount\minutes  \newcount\temp \newcount\pmhours
\hours = \time
\divide\hours by 60
\temp = \hours
\multiply\temp by 60
\minutes = \time
\advance\minutes by -\temp
\def\hour{\the\hours}
\def\minute{\ifnum\minutes<10 0\the\minutes
            \else\the\minutes\fi}
\def\clock{
\ifnum\hours=0 12:\minute\ AM
\else\ifnum\hours<12 \hour:\minute\ AM
      \else\ifnum\hours=12 12:\minute\ PM
            \else\ifnum\hours>12
                 \pmhours=\hours
                 \advance\pmhours by -12
                 \the\pmhours:\minute\ PM
                 \fi
            \fi
      \fi
\fi
}

\def\monthname{\relax\ifcase\month 0/\or January\or February\or
   March\or April\or May\or June\or July\or August\or September\or
   October\or November\or December\else\number\month/\fi}

\def\bold#1{\setbox0=\hbox{$#1$}%
     \kern-.025em\copy0\kern-\wd0
     \kern.05em\copy0\kern-\wd0
     \kern-.025em\raise.0433em\box0 }

\newcommand{\Mh}{\ensuremath{M_h}}
\newcommand{\MA}{\ensuremath{M_A}}
\newcommand{\mst}{m_{\tilde t}}
\newcommand{\tb}{\ensuremath{\tan\beta}}
\newcommand{\msqd}{m_{\widetilde{q_3}}}
\newcommand{\At}{A_t}

\newcommand{\neu}[1]{\tilde \chi^0_{#1}}
\newcommand{\mneu}[1]{m_{\tilde \chi^0_{#1}}}
\def\order#1{\ensuremath{{\cal O}(#1)}}
\newcommand{\DRbar}{\ensuremath{\overline{\mathrm{DR}}}}

\newcommand{\als}{\alpha_s}
\newcommand{\alt}{\alpha_t}
\newcommand{\fh}{{\tt FeynHiggs}}

\def\beq{\begin{equation}}
\def\eeq{\end{equation}}

\def\ga{\mathrel{\raise.3ex\hbox{$>$\kern-.75em\lower1ex\hbox{$\sim$}}}}
\def\la{\mathrel{\raise.3ex\hbox{$<$\kern-.75em\lower1ex\hbox{$\sim$}}}}
\def\gyr{{\rm \, G\kern-0.125em yr}}

\def\gappeq{\mathrel{\rlap {\raise.5ex\hbox{$>$}}
{\lower.5ex\hbox{$\sim$}}}}
\def\lappeq{\mathrel{\rlap{\raise.5ex\hbox{$<$}}
{\lower.5ex\hbox{$\sim$}}}}
\def\Toprel#1\over#2{\mathrel{\mathop{#2}\limits^{#1}}}

\def\m12{m_{1\!/2}}

\def\msl{m_{\tilde{\ell}_1}}

\def\mt{m_{t}}

\def\msqfs{m_{\widetilde{q}_{12}}}
\def\msqt{m_{\widetilde{q}_3}}
\def\msl{m_{\widetilde{l}}}
\newcommand{\gmt}{\ensuremath{(g-2)_\mu}}
\newcommand{\br}{{\rm BR}}
\newcommand{\bsg}{\ensuremath{\br(b \to s \gamma)}}
\newcommand{\bmm}{\ensuremath{\br(B_s \to \mu^+\mu^-)}~}
\newcommand{\gsim}{\lower.7ex\hbox{$\;\stackrel{\textstyle>}{\sim}\;$}}
\newcommand{\lsim}{\lower.7ex\hbox{$\;\stackrel{\textstyle<}{\sim}\;$}}

\newcommand{\tev}{\,\, \mathrm{TeV}}
\newcommand{\gev}{\,\, \mathrm{GeV}}

\newcommand{\ETslash}{\ensuremath{/ \hspace{-.7em} E_T}}

\graphicspath{{figs/}}

\begin{document}
\begin{titlepage}
\pagestyle{empty}
\begin{center}
{\small KCL-PH-TH/2013-41, LCTS/2013-28, CERN-PH-TH/2013-294, MPP-2013-313 \\
DESY~13-249, FR-PHENO-2013-019, UMN--TH--3315/13, FTPI--MINN--13/42, SLAC-PUB-15855}
\end{center}
\vskip 0.15in
\begin{center}
{\large {\bf Implications of Improved Higgs Mass Calculations \\ for
 Supersymmetric Models}} \\
\end{center}
\begin{center}
\vskip 0.20in
{\bf O.~Buchmueller}$^1$, {\bf M.J.~Dolan}$^2$,
{\bf J.~Ellis}$^3$, {\bf T.~Hahn}$^4$, 
{\bf S.~Heinemeyer}$^5$, 
{\bf W.~Hollik}$^4$, \\
{\bf J.~Marrouche}$^{1}$, {\bf K.~A.~Olive}$^{6}$, 
{\bf H.~Rzehak}$^7$, {\bf K.~de~Vries}$^{1}$
and {\bf G.~Weiglein}$^{8}$
\vskip 0.2in
{\it
$^1${High Energy Physics Group, Blackett Lab., Imperial College, Prince Consort Road, London SW7 2AZ, UK}\\
$^2${Theory Group, SLAC National Accelerator Lab.,
2575 Sand Hill Road, Menlo Park, \\ CA 94025-7090, USA}\\
$^3${Theoretical Particle Physics and Cosmology Group, Dept.\ of Physics, 
King's College London, London WC2R 2LS, UK;\\
Theory Division, CERN, CH-1211 Geneva 23, Switzerland}\\
$^4${Max-Planck-Institut f{\"u}r Physik, F{\"o}hringer Ring 6, D-80805 Munich, Germany}\\
$^5${Instituto de F{\'i}sica de Cantabria (CSIC-UC), E-39005 Santander, Spain}\\
$^6${William\,I.\,Fine\,Theoretical\,Physics\,Institute,\,%
School of Physics and Astronomy, 
Univ.\,of\,Minnesota,\,%
Minneapolis,\,MN\,55455,\,USA}\\
$^7${Albert-Ludwigs-Universit\"at Freiburg, 
Physikalisches Institut, D--79104 Freiburg, Germany}\\
$^8${DESY, Notkestrasse 85, DÐ22607 Hamburg, Germany}}\\
\vskip 0.2in 
{\bf Abstract}
\vskip 0.2in 
\end{center}
\noindent

We discuss the allowed parameter spaces of supersymmetric scenarios in
light of improved Higgs mass predictions provided by 
{\tt FeynHiggs~2.10.0}. The Higgs mass predictions combine Feynman-diagrammatic
results with a resummation of leading and subleading
logarithmic corrections from the stop/top sector, which yield a significant improvement in the region of
large stop masses.
Scans in the pMSSM parameter space show that, for given values of the soft
supersymmetry-breaking parameters, the new logarithmic contributions beyond the two-loop order 
implemented in {\tt FeynHiggs} tend to give larger
values of the light CP-even Higgs mass, \Mh, in the region of large stop masses than previous
predictions that were based on a fixed-order Feynman-diagrammatic result,
though the differences are generally consistent with the previous
estimates of theoretical uncertainties.
We re-analyze the parameter spaces of the
CMSSM, NUHM1 and NUHM2, taking into account also the constraints from
CMS and LHCb measurements of \bmm and ATLAS
searches for \ETslash\ events using 20/fb of LHC
data at 8~TeV. Within the CMSSM, the Higgs mass constraint disfavours 
$\tb \lsim 10$, though not in the NUHM1 or NUHM2.

\end{titlepage}
\baselineskip=16pt

\section{Introduction}

The ATLAS and CMS experiments did not discover supersymmetry
(SUSY) during the first,
low-energy LHC run at 7 and 8~TeV. However, an optimist may consider that the
headline discovery of a Higgs boson weighing $\sim 126 \gev$~\cite{lhch}
has provided two 
additional pieces of indirect, circumstantial evidence for SUSY, beyond the many
previous motivations. One piece of circumstantial evidence is provided by the Higgs mass,
which falls within the range $\lsim 135 \gev$ calculated in the minimal
SUSY extension of the Standard Model 
(MSSM) for masses of the SUSY particles around 
1~TeV~\cite{mh,mh2loop,mhiggsAEC,susycompare}. The other 
piece of circumstantial evidence is provided by measurements of Higgs couplings, which
do not display any significant deviations from Standard Model (SM)
predictions at the present level of experimental accuracy. This disfavours
some composite models but is consistent with the predictions of simplified SUSY
models such as the constrained MSSM (CMSSM) \cite{cmssm} with universal
input soft SUSY-breaking masses $m_0$ for scalars, $m_{1/2}$ for
fermions as well as $A_0$, the soft SUSY-breaking trilinear coupling, 
and NUHM models that have non-universal soft
SUSY-breaking contributions to Higgs supermultiplet masses: see~\cite{nuhm2,nuhm1},
and~\cite{AbdusSalam:2011fc} for a review.

That said, the absence of SUSY in the first LHC run and the fact that
the Higgs mass is in the upper part of the MSSM range both suggest,
within simple models such as the CMSSM and NUHM (see,
e.g.,\cite{eo6,elos}) as well as in the pMSSM, that
the SUSY particle mass scale may be larger than had been suggested prior to
the LHC, on the basis of fine-tuning arguments and in order to explain
the discrepancy between calculations of \gmt\ within the SM and the
experimental  measurement \cite{g-2}. A relatively large SUSY particle
mass scale also makes it easier to reconcile SUSY with the experimental
measurement of \bmm\ \cite{bmm}, particularly if $\tb$ (the ratio of
SUSY Higgs vacuum expectation values, v.e.v.s) is large. 

The mathematical connection between the Higgs mass and the SUSY
particle spectrum 
is provided by calculations of the lightest SUSY Higgs mass \Mh\ in terms of
the SUSY particle spectrum \cite{mh,mh2loop,mhiggsAEC,susycompare}:
see \cite{MSSMreviews} for reviews.
As is well-known, one-loop radiative corrections allow
\Mh\ to exceed $M_Z$ by an amount that is logarithmically sensitive to such
input parameters as the top squark masses $m_{\tilde{t}}$ in the pMSSM, or the universal $m_{1/2}$ and  $m_0$ masses
in the CMSSM and NUHM.
Inverting this calculation, the inferred values of $m_{\tilde{t}}$, or $m_{1/2}, m_0$ and $A_0$ are
exponentially sensitive to the measured value of \Mh. For this reason, it is
essential to make available and use the most accurate calculations of \Mh\
within the MSSM, and to keep track of the unavoidable theoretical uncertainties
in these calculations due to unknown higher-order corrections, which are now
larger than the experimental measurement error.

Several codes to calculate \Mh\ are available~\cite{softsusy,spheno,suspect,cpsuperh,h3m}.
In terms of low-energy parameters, the most advanced calculation is
provided by {\tt FeynHiggs}~\cite{FeynHiggs,mhiggsAEC}.
The differences between the codes are in the few GeV range
for relatively light SUSY spectra, but may become larger for higher third family squark masses and 
values of $m_{1/2}, m_0$ and $A_0$. This is particularly evident in the phenomenological MSSM (pMSSM), 
where the soft supersymmetry-breaking inputs to the SUSY spectrum codes are specified at a low scale, 
close to the physical masses of the supersymmetric particles.

In this paper we revisit the constraints on the CMSSM and NUHM parameter
spaces imposed by the Higgs mass measurement using the significantly improved {\tt 2.10.0}
version of the {\tt FeynHiggs} code \cite{FeynHiggs,newFH} that has recently been released. We situate our discussion
in the context of a comparison between this and the earlier version 
{\tt FeynHiggs~2.8.6}, 
which has often been used in phenomenological studies of SUSY parameter
spaces (e.g., in \cite{mc8}), as well as with {\tt SOFTSUSY~3.3.9}~\cite{softsusy}. 
We also discuss the implications for constraints on 
SUSY model parameters. Updating previous related analyses \cite{eo6,elos},
we also take into account the complementary constraint on the CMSSM and 
NUHM parameter spaces imposed by the recent experimental measurement of \bmm,
and we incorporate the 95\%~CL limit on $m_{1/2}$ and $m_0$ established
within the CMSSM by ATLAS following searches for missing transverse
energy, ~\ETslash,
events using 20/fb of LHC data at 8~TeV \cite{ATLASsusy}.

The layout of this paper is as follows. In Section~2 we first summarize the main improvements 
between the results implemented in {\tt FeynHiggs~2.8.6} and {\tt 2.10.0}, and
then present some illustrative results in the pMSSM, discussing the numerical
differences between calculations made using {\tt FeynHiggs} versions
{\tt 2.8.6} and {\tt 2.10.0}. 
We then display in Section~3 some representative parameter planes in the CMSSM,
NUHM1 and NUHM2, discussing the interplay between the different experimental constraints
including \bmm as well as \Mh. Section~4 contains a discussion
of the variations between the predictions of \Mh\ made in global fits to CMSSM and NUHM1
model parameters using different versions of {\tt FeynHiggs}
and {\tt SOFTSUSY}. Finally, Section~5 summarizes
our conclusions.


\section{Comparisons of Higgs Mass Calculations within the General MSSM}

\subsection{The improved Higgs Mass Calculation in {\tt FeynHiggs~2.10.0}}

The evaluation of Higgs boson masses in the MSSM, in particular of
the mass of the lightest Higgs boson, \Mh, has recently been improved
for larger values of the scalar top mass scale. This new evaluation has
been implemented in the code {\tt FeynHiggs~2.10.0}, whose details
can be found in~\cite{newFH}. Here we just summarize some salient
points.

The code {\tt FeynHiggs} provides predictions for the
masses, couplings and decay properties of the
MSSM Higgs bosons
at the highest currently available level of accuracy 
as well as approximations for LHC
production cross sections
(for MSSM Higgs decays see also~\cite{yr3} and
references therein). 
The evaluation of Higgs boson masses within {\tt FeynHiggs} 
is based on a Feynman-diagrammatic calculation of the
Higgs boson self-energies. By finding the higher-order corrected poles
of the propagator matrix, the loop-corrected Higgs boson masses are
obtained.

The principal focus of the improvements in {\tt FeynHiggs~2.10.0} has been to attain
greater accuracy for large stop masses. The versions of {\tt FeynHiggs} as
used, e.g., previously in~\cite{mc8} included the full one-loop and
the leading and subleading two-loop corrections to the Higgs boson self-energies
(and thus to \Mh). The new version, {\tt FeynHiggs~2.10.0}~\cite{newFH}, which is used for the evaluations
here, contains in addition a resummation of the leading and next-to-leading logarithms of type
$\log(\mst/\mt)$ in all orders of perturbation theory,
which yields reliable results for $\mst, \MA \gg M_Z$.
To this end the two-loop Renormaliz\-ation-Group Equations
(RGEs)~\cite{SM2LRGE} have been solved, taking into account the one-loop threshold
corrections to the quartic coupling at the SUSY scale:
see~\cite{hep-ph/0001002} and references therein. 
In this way at $n$-loop order the terms 
\begin{align}
\sim \; \log^n (\mst/\mt), \quad \sim \; \log^{n-1}(\mst/\mt)
\end{align}
are taken into account.
The resummed logarithms, which are calculated in the $\overline{{\rm MS}}$
scheme for the scalar top sector, are matched to the one- and two-loop
corrections, where the on-shell scheme had been used for the scalar top
sector. The first main difference between {\tt FeynHiggs~2.10.0} and previous
versions occurs at three-loop order.  As we shall see,
{\tt FeynHiggs~2.10.0} yields a larger estimate of \Mh\ for stop masses
in the multi-TeV range, and a correspondingly improved estimate of the
theoretical uncertainty, as discussed in~\cite{newFH}.
The improved estimate of the uncertainties arising from
corrections beyond two-loop order in the top/stop sector is adjusted
such that the impact of replacing the running top-quark mass by the pole
mass (see \cite{mhiggsAEC}) is evaluated only for the non-logarithmic
corrections rather than for the full two-loop contributions implemented
in \fh.

Other codes such as \texttt{SoftSusy}~\cite{softsusy},
\texttt{SPheno}~\cite{spheno} and \texttt{SuSpect}~\cite{suspect}  
implement a calculation of the Higgs masses based on a 
$\overline{{\rm DR}}$ renormalization of the scalar quark sector~\footnote{Since
the differences between the on-shell and
$\overline{{\rm DR}}$ renormalization in the scalar quark sector are  
formally of higher order, comparisons can be used to assess the uncertainties in the
predictions of the Higgs mass.}.
These codes contain the full one-loop corrections to the MSSM Higgs masses and
implement the most important two-loop corrections. In particular, 
\texttt{SoftSusy} contains the $\mathcal{O}(\alpha_t^2)$, 
$\mathcal{O}(\alpha_b\alpha_\tau)$, $\mathcal{O}(\alpha_b^2)$, 
$\mathcal{O}(\alpha_b\alpha_s)$, $\mathcal{O}(\alpha_t \alpha_s)$, 
$\mathcal{O}(\alpha_\tau^2)$  and $\mathcal{O}(\alpha_t \alpha_b)$ corrections
of~\cite{mh2loop,susycompare} evaluated at zero external momentum  
for the neutral Higgs masses. 
These codes do not contain the additional resummed higher-order terms
included in {\tt FeynHiggs~2.10.0}.
We return in Section~\ref{sec:mastercode} to a comparison between
\texttt{SoftSusy3.3.9} and $\texttt{FeynHiggs2.10.0}$. 

More recently a calculation of \Mh\ taking into account leading three-loop corrections of
\order{\alt\als^2} 
has became available, based on a \DRbar\ or a ``hybrid'' renormalisation scheme
for the scalar top sector, where the numerical
evaluation depends on the various SUSY mass hierarchies, 
resulting in the code {\tt H3m}~\cite{h3m}, which
adds the three-loop corrections to the \fh\ result.
A brief comparison between \fh\ and {\tt H3m} can be found in~\cite{FKPS,newFH}.

A numerical analysis in the CMSSM including leading three-loop
corrections to $\Mh$ (with the code {\tt H3m}) was presented
in~\cite{FKPS}. It was shown that the leading three-loop terms can have a
strong impact on the interpretation of the measured Higgs mass value in the
CMSSM. Here, with the new version of {\tt FeynHiggs}, we go beyond this
analysis by including (formally) subleading three-loop corrections as well as
a resummation to all orders of the leading and next-to-leading logarithmic contributions to $\Mh$, see
above.


\subsection{Comparing the improved Higgs Mass Calculation in\\ 
{\tt FeynHiggs~2.10.0} with {\tt FeynHiggs~2.8.6}}
\label{sec:pmssm}

In the following we examine the effect of 
including the resummation of leading and subleading logarithmic
corrections from the (scalar) top sector in the pMSSM.
We compare the new {\tt FeynHiggs} version {\tt 2.10.0} with a
previous one, {\tt 2.8.6}, where the only relevant difference in the
Higgs mass calculation between the two codes consists of the
aforementioned resummation effects.
(A comparison including {\tt SOFTSUSY} can be found in
Sect.~\ref{sec:mastercode}.) 
These corrections are most sensitive to the soft SUSY-breaking parameters
  in the stop sector, $\msqd$ in the diagonal entry (which we assume
  here to be equal for left- and right-handed stops) and the
trilinear coupling $\At$. 
To have direct control over these two parameters, we consider a 10-parameter 
incarnation of the MSSM, denoted as the pMSSM10. 
In the pMSSM10 we set the squark masses of the first two generations  to
a common value $\msqfs$, the third-generation squark mass parameters to a
different value $\msqt$,  
the slepton masses to $\msl$, and the trilinear couplings $\At = A_b =
A_{\tau} = A$. 
The remaining parameters of the pMSSM10 are the soft SUSY-breaking
parameters in the gaugino sectors, $M_1,~M_2,~M_3$, the Higgs mixing
parameter $\mu$, the CP-odd Higgs mass scale $\MA$ as well as $\tb$.

We generate 1000 random sets of the 8 parameters
$\msqfs~\msl,~M_1,~M_2,~M_3,~\tb,~\mu$ and $\MA$, 
without regard to the experimental constraints. 
For each of these sets we vary $\msqt=0.5,~1,~2,~3,~4$ and 5~TeV and $A/\msqt=0, \pm1.0, \pm2.0, \pm2.4$,
and calculate the corresponding spectra using {\tt SOFTSUSY-3.3.9}. 
Using these spectra, we calculate $\Mh$ with {\tt FeynHiggs~2.8.6} and {\tt FeynHiggs~2.10.0}. 
We stress that the pMSSM10 spectra are only meant to illustrate the size of the
corrections as a function of $\msqd$ and the trilinear coupling $A$,
and do not necessarily correspond to phenomenologically interesting
regions of parameter space.

\begin{figure}[hbtp!]
\centerline{
\includegraphics[height=20cm]{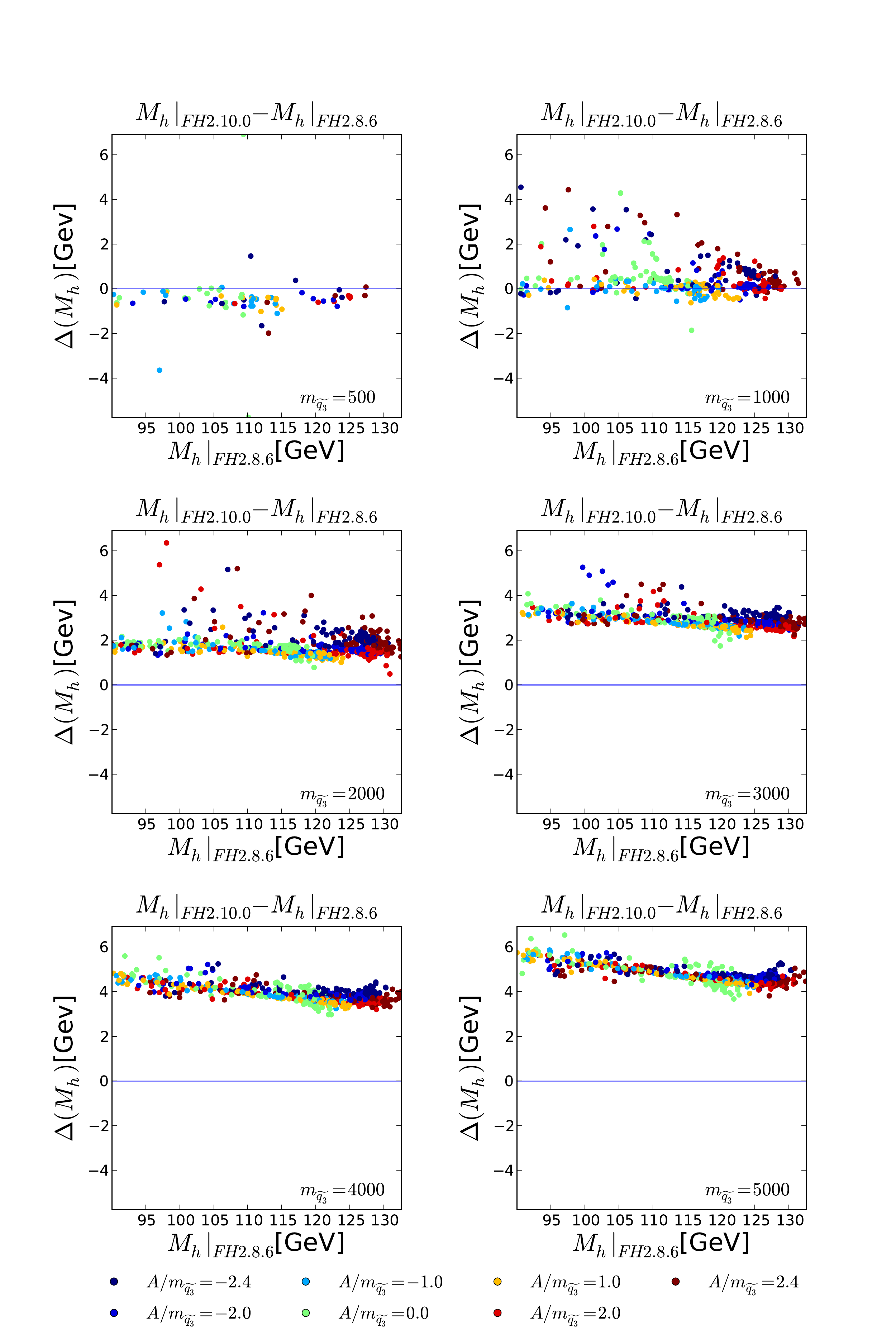}
}
\caption{\label{fig:2862100}\it
The differences between \Mh\ calculated using {\tt FeynHiggs~2.10.0} and {\tt FeynHiggs~2.8.6},
as a function of the {\tt FeynHiggs~2.8.6} value, for third-generation squark masses $\msqt =$
0.5~TeV (upper left), 1~TeV (upper right), 2~TeV (middle left), 
3~TeV (middle right),
4~TeV (lower left) and 5~TeV (lower right). }
\end{figure}

The sizes of the corrections from the (scalar) top sector are given by the 
differences $(\Mh|_{{\rm FH}2.10.0}-\Mh|_{{\rm FH}2.8.6})$ 
shown in Fig.~\ref{fig:2862100} as functions of $\Mh|_{{\rm FH}2.8.6}$.
The different panels in this figure correspond to the different third-generation squark masses $\msqt =$
0.5~TeV (upper left), 1~TeV (upper right), 
2~TeV (middle left), 3~TeV (middle right),
4~TeV (lower left) and 5~TeV (lower right), whereas
 the colours dark blue, blue, light blue, light green, orange, red and dark red 
correspond to $A/\msqt=-2.4,~-2.0,~-1.0,~0.0,~1.0,~2.0,~2.4$ respectively.
At low stop masses of around 500~GeV we see that
the resummation corrections are ${\cal O}(0.5)$~GeV, 
whereas with increasing stop masses they may become as large as 5~GeV.
The dependence on $A/\msqt$ is less significant.
We also note that, for similar values of $\msqt$, the resummation corrections tend to be smaller
for models yielding $\Mh \sim 125 \gev$ than for models yielding smaller
values of $\Mh$.

The latter effect is related to the (random) choice of $\MA$ and
  $\tb$, with lower $\Mh$ values corresponding to lower $\MA$ and smaller
  $\tb$. If the $\Mh$ value without resummed corrections, i.e., from 
{\tt FeynHiggs~2.8.6}, is smaller, the newly added correction, which is
independent of $\MA$ and $\tb$ has a larger effect.
We should furthermore mention that the size of the resummed correction
stays (mostly) within the previously-predicted estimate for the
theoretical uncertainties due to missing higher-order
corrections. Consequently, a point in the MSSM parameter space that has a
Higgs mass value of, for instance,  $125 \gev$ as evaluated by 
{\tt FeynHiggs~2.10.0}, should not have been excluded on the basis of a
lower $\Mh$ as evaluated using {\tt FeynHiggs~2.8.6}. However, the
parallel reduction in the theory uncertainty in {\tt FeynHiggs~2.10.0} leads to a more precise
restriction on the allowed MSSM parameter space.

\section{Examples of CMSSM and NUHM Parameter Planes}

In our exploration of the {\tt FeynHiggs~2.10.0} results for \Mh, we
discuss their interplay with other experimental constraints, notably
\bmm\ and the ATLAS search for \ETslash\ events with 20/fb of data at
8~TeV.  In this section, results were produced using SSARD \cite{SSARD} coupled
to {\tt FeynHiggs}. These results update those in \cite{eo6} for the CMSSM and \cite{elos}
for the NUHM.
In the case of the CMSSM, we consider several $(m_{1/2}, m_0)$
planes for fixed values of $\tb$ and $A_0/m_0$, all with $\mu > 0$. In
the NUHM1 model we also display two $(m_{1/2}, m_0)$ planes for fixed
values of $\tb$ and $A_0/m_0$, one with fixed $\mu = 500 \gev$ and one
with fixed $\MA = 1000 \gev$, and two $(\mu, m_0)$ planes with fixed
$\tb, m_{1/2}$ and $A_0/m_0$. In the NUHM2 we display two $(\mu, \MA)$
planes with fixed $\tb, m_{1/2}, m_0$ and $A_0/m_0$. We also present one
example of a $(m_{1/2}, m_0)$ plane in the minimal supergravity (mSUGRA)
model, in which the electroweak vacuum conditions fix $\tb$ as a
function of $m_{1/2}, m_0$ and $A_0$. 

We adopt the following conventions in all these figures.  
Regions where the LSP is charged are shaded brown, those where there is
no consistent electroweak vacuum are shaded mauve, regions excluded by
\bsg\ measurements at the 2-$\sigma$ level are shaded green, 
those favoured by the SUSY interpretation
of \gmt\ are shaded pink, with lines indicating the $\pm 1\sigma$ (dashed) and $\pm 2\sigma$ ranges
(solid), and strips with an LSP density appropriate to make up all the cold
dark matter are shaded dark blue. 
For reasons of visibility, we shade
strips where $0.06 < \Omega_\chi h^2 < 0.2$, but when we quote ranges of
consistency we require that the relic density satisfies the more restrictive relic density bound $0.115 < \Omega_\chi h^2 < 0.125$ \cite{planck}.
The 95\%~CL limit from the ATLAS \ETslash\ search is shown as a continuous
purple contour~\footnote{The ATLAS \ETslash\ limit was quoted for the CMSSM
with the choices $\tb = 30$ and $A_0/m_0 = 2$, but a previous
study~\cite{mc8} showed that such a contour is essentially independent
of both $\tb$ and $A_0/m_0$, as well as the amount of non-universality
in NUHM models.}%
, and the 68 and 95\%~CL
limits from the CMS and LHCb measurements of \bmm\  
are shown as continuous green contours.
Finally, the labelled continuous black lines are contours of \Mh\
calculated with {\tt FeynHiggs~2.10.0}, and the dash-dotted red lines
are contours of \Mh\ calculated with {\tt FeynHiggs~2.8.6}
(as used, e.g., in~\cite{eo6,elos,mc8}), which we use
for comparison.


\subsection{The CMSSM}

\begin{figure}
\vspace{-2cm}
\begin{center}
\begin{tabular}{c c}
\includegraphics[height=8cm]{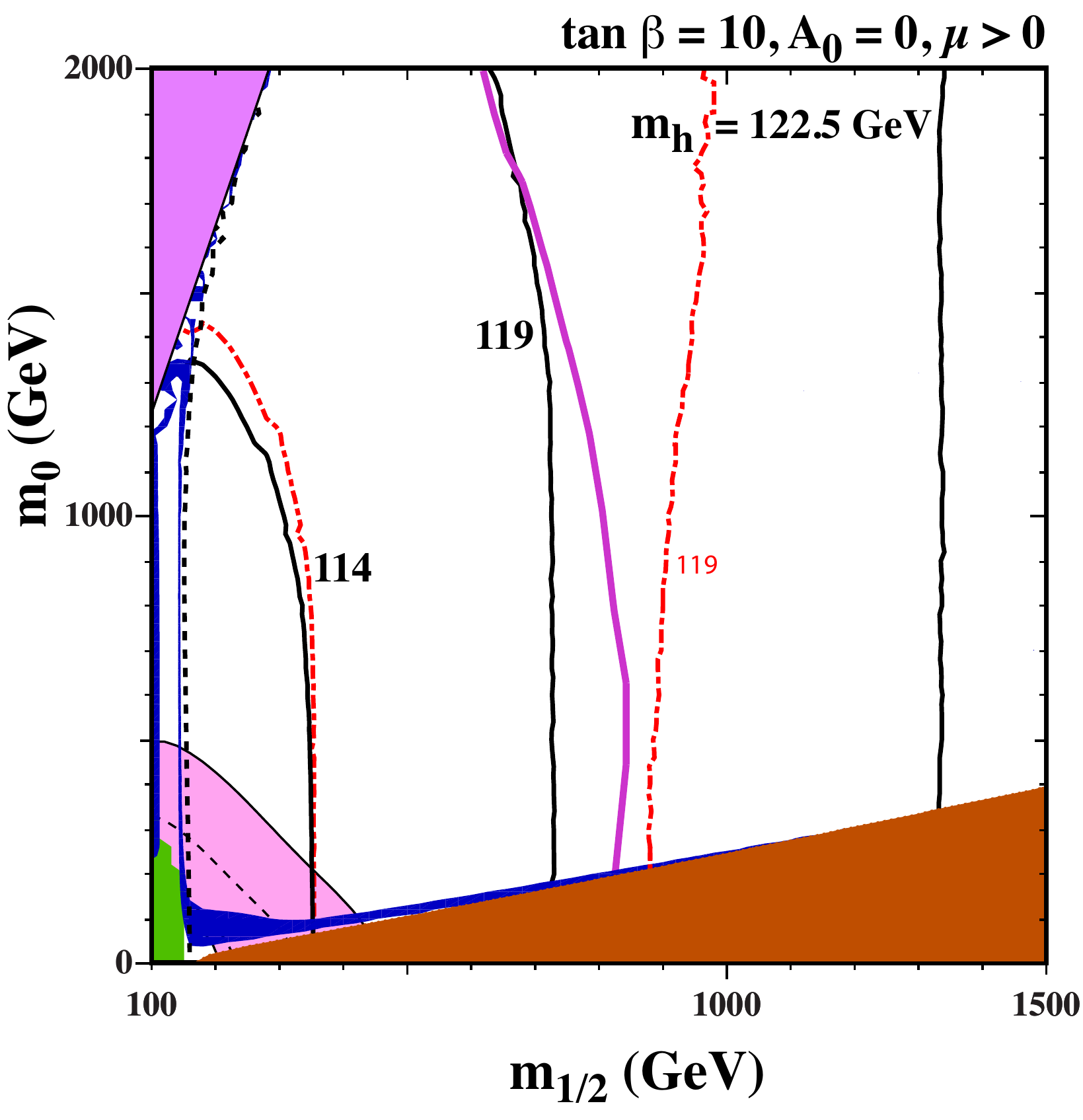} & 
\includegraphics[height=8cm]{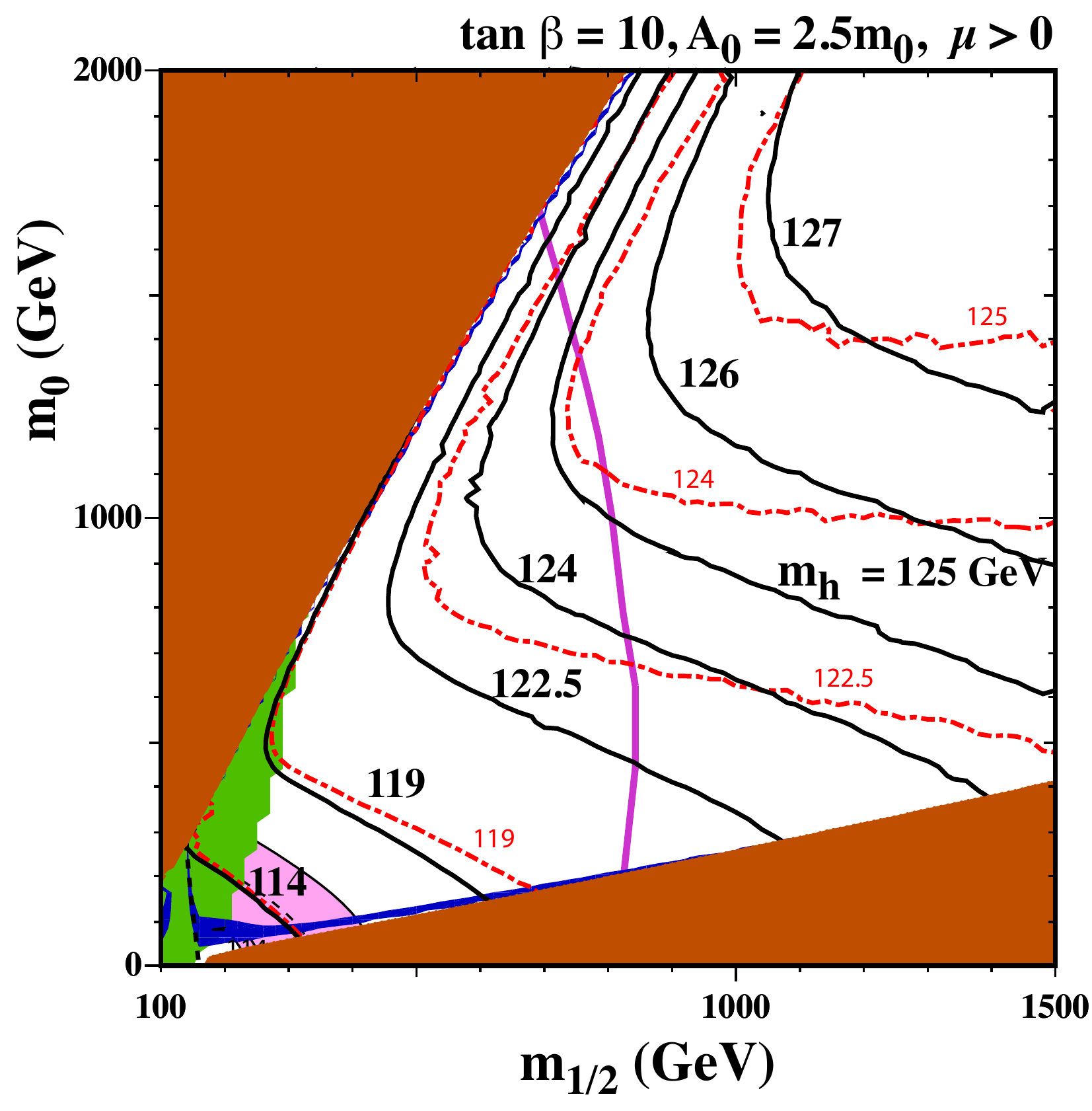} \\
\end{tabular}
\end{center}   
\begin{center}
\begin{tabular}{c c}
\includegraphics[height=8cm]{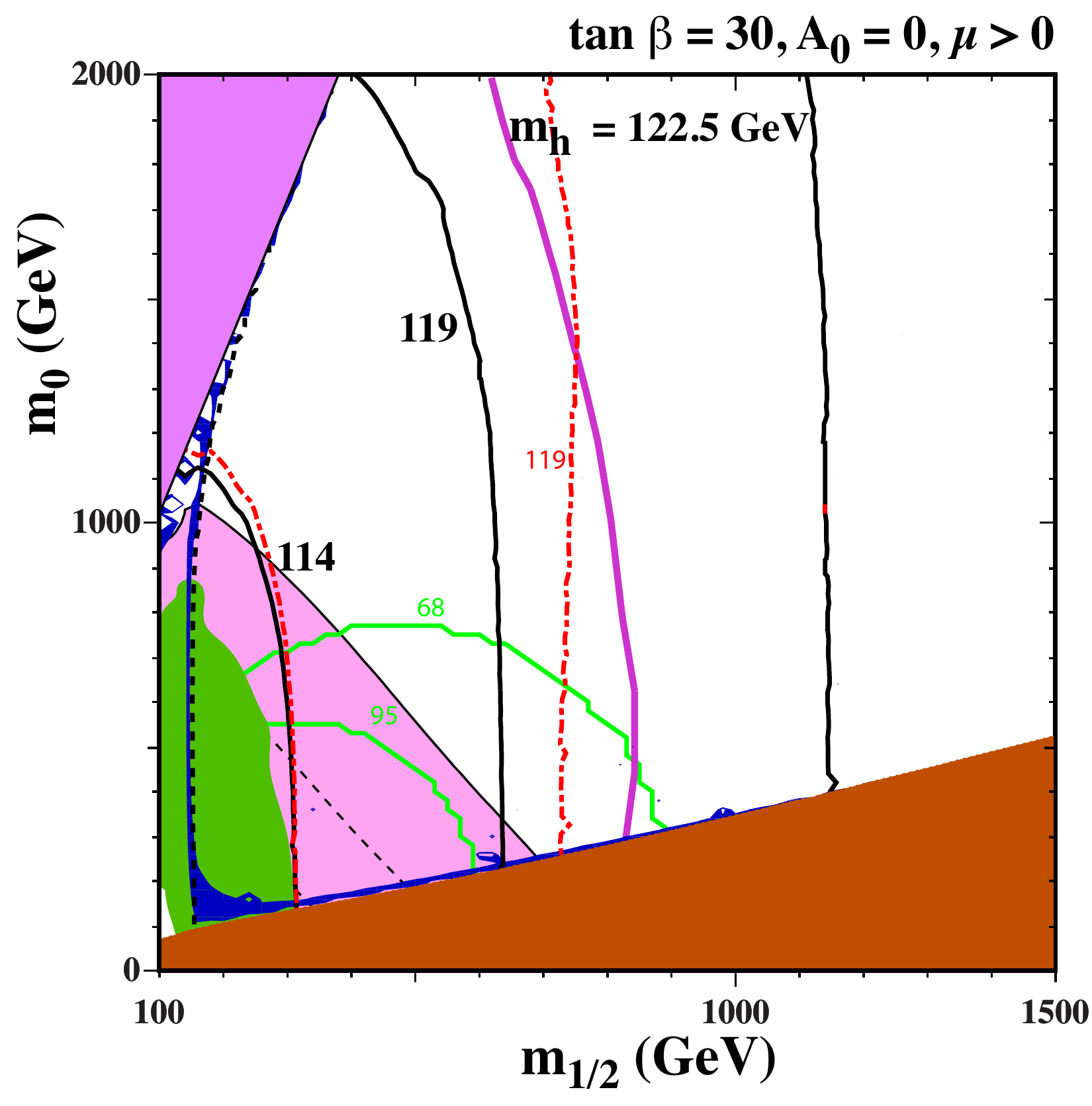} & 
\includegraphics[height=8cm]{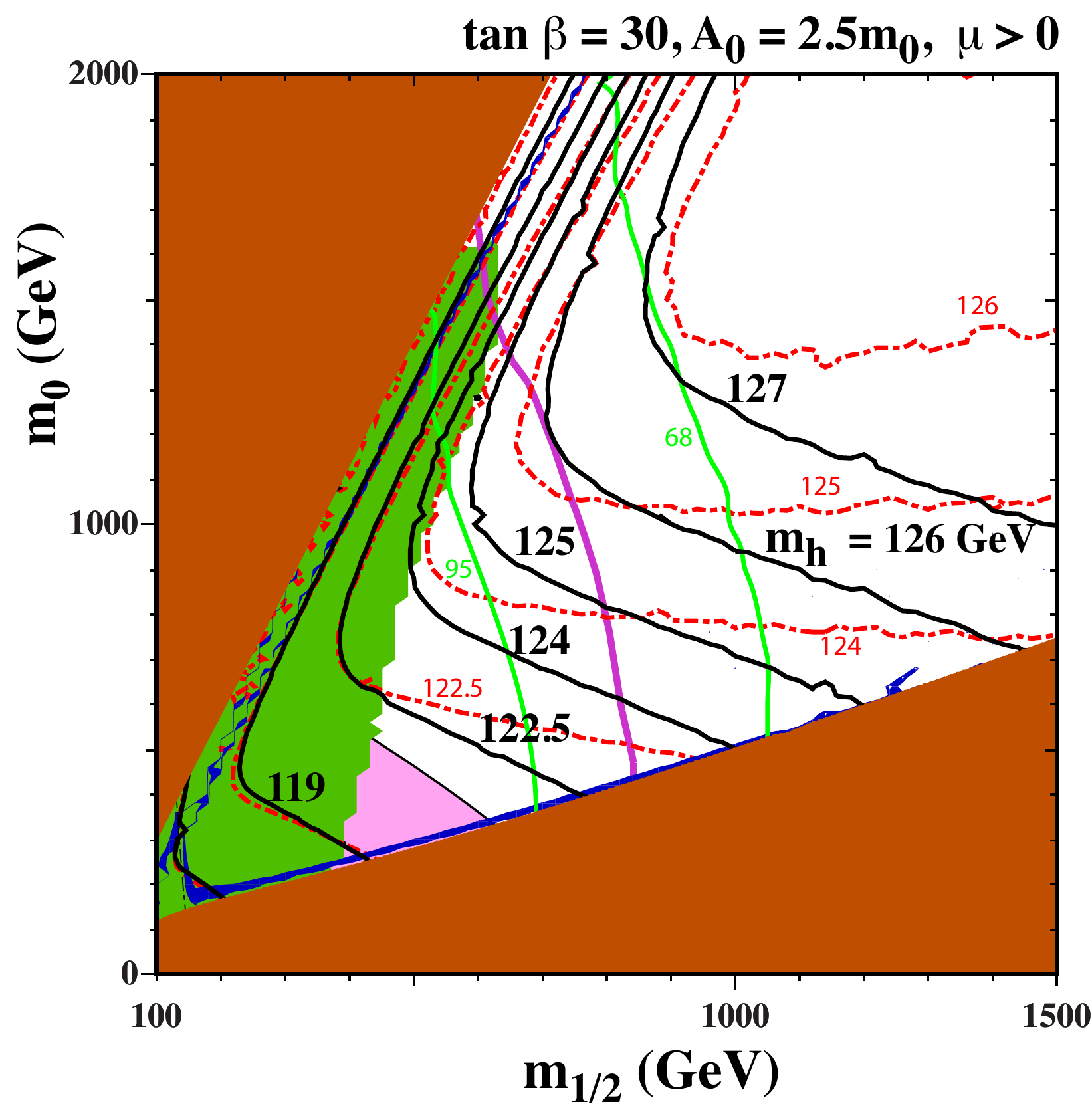} \\
\end{tabular}
\end{center}   
\caption{\label{fig:1030}\it
The allowed regions in the $(m_{1/2}, m_0)$ planes for $\tb = 10$ and $A_0 = 0$
(upper left), $\tb = 10$ and $A_0 = 2.5 m_0$ (upper right), $\tb = 30$ and $A_0 = 0$
(lower left) and $\tb = 30$ and $A_0 = 2.5 m_0$ (lower right). The line styles and shadings are
described in the text. The section of the dark blue coannihilation strip in the lower right panel
in the range $m_{1/2} \in (840, 1050) \gev$
is compatible with the constraints from \bmm\ (green line) and the
ATLAS 20/fb \ETslash\ search (purple line), as well as with the LHC \Mh\ measurement.
Better
consistency with all the constraints (except \gmt) 
is found if the improved {\tt FeynHiggs~2.10.0} code is used, 
for $\tb = 30$ and $A_0 = 2.5 m_0$.
}
\end{figure}

Fig.~\ref{fig:1030} displays four examples of $(m_{1/2}, m_0)$ planes for relatively low values
of $\tb$. We see in the upper left panel for $\tb = 10$ and $A_0 = 0$ that the contour for $\Mh = 114 \gev$
(the lower limit set by the LEP experiments) changes very little between {\tt FeynHiggs~2.8.6} and {\tt 2.10.0}, whereas that for 119~GeV is
shifted by $\Delta m_{1/2} \sim - 150 \gev$ in the region of the stau-coannihilation strip at low $m_0$. The ATLAS 20/fb \ETslash\
limit on $m_{1/2}$ excludes robustly a SUSY solution to
the \gmt\ discrepancy in this particular CMSSM scenario, but neither $b \to s \gamma$ nor
$B_s \to \mu^+ \mu^-$ has any impact on the allowed section of the dark matter strip, which
extends to $m_{1/2} \sim 900 \gev$ in this case. However, none
of it is compatible with the measured value of \Mh, even 
with the higher value and the correspondingly smaller theory
  uncertainty as evaluated by
{\tt FeynHiggs~2.10.0} which is about $\pm 0.8 \gev$ near the
endpoint of the strip. 
There is a mauve region at small $m_{1/2}$ and large $m_0$ where the electroweak vacuum
conditions cannot be satisfied, adjacent to which there is a portion of a focus-point strip, excluded by the
ATLAS \ETslash\ search, where \Mh\ is smaller than the measured value.

In the upper right panel of Fig.~\ref{fig:1030}, which displays the case
$\tb = 10$ and $A_0 = 2.5 m_0$, we see that the {\tt FeynHiggs~2.10.0} $\Mh = 119 \gev$
contour intersects the stau-coannihilation strip when $m_{1/2} \sim 600 \gev$ (a shift of less than 100 GeV in $m_{1/2}$
compared to {\tt FeynHiggs~2.8.6}) and the tip of the strip
corresponds to $\Mh \sim 122 \gev$. The experimental value of $\Mh$ lies somewhat outside the
range around this value that is allowed by the uncertainty estimated
in {\tt FeynHiggs~2.10.0}, which is about $1.0 \gev$ at this
point. Consequently, although 
the use of {\tt FeynHiggs~2.10.0} reduces significantly the tension with the measurement of
\Mh\ for this value of $\tb$ in the CMSSM, it seems that this model requires a
larger value of $\tb$.

We note in this case the appearance of a brown region in the upper left part of the plane, 
where the lighter scalar top is the LSP (or tachyonic), with an
adjacent stop-coannihilation strip. We find $\Mh < 122 \gev$ 
in the displayed section of the strip where $m_0 < 2000 \gev$, but larger values of \Mh\ can be
found at larger $m_0$, which may be compatible with the LHC measurement, within the uncertainties.
For example, at $m_{1/2} = 1500$~GeV, the stop-coannihilation strip is found at $m_0 \simeq 3450 \gev$
and the Higgs mass there computed with {\tt FeynHiggs~2.10.0} is $\Mh \simeq 125 \gev$, 
substantially higher than the value of 121 GeV found in {\tt FeynHiggs~2.8.6}, though with a
larger uncertainty of 2~GeV. 

The lower left panel of Fig.~\ref{fig:1030} displays the $(m_{1/2}, m_0)$ plane for 
$\tb = 30$ and $A_0 = 0$. Compared with the $\tb = 10, A_0 = 0$ case,
the Higgs mass contours are similar, though shifted somewhat to lower $m_{1/2}$. 
The focus-point region is found at slightly larger $m_{1/2}$ but
is not very different from the $\tb = 10$ case.
We note also the appearance of the (green) 68\% and 95\%~CL constraints from 
\bmm, though the constraints from the ATLAS \ETslash\ search and (particularly) \Mh\ are 
more important. Although the stau-coannihilation strip extends to slightly
higher values of $m_{1/2} \sim 1000 \gev$ when $A_0 = 0$, the Higgs mass at the endpoint is still
only $122 \pm 0.8 \gev$. It is well known that the calculated value of \Mh\ increases with the value of 
$A_0$, and compatibility with the LHC measurement for this value of $\tb$ requires a larger value
of $A_0$.

Accordingly, in the lower right panel of Fig.~\ref{fig:1030} we show the case of 
$\tb = 30$ and $A_0 = 2.5 m_0$.
As expected, the situation along the
stau-coannihilation strip is much more favourable for \Mh. At the end point of the 
stau-coannihilation strip, which is now at about 
$m_{1/2} \simeq 1250 \gev$, according to the improved {\tt FeynHiggs~2.10.0} calculation the Higgs mass is 
$\Mh \simeq 125.2 \pm 1.1 \gev$, quite consistent with LHC measurement,
whereas the previous version of {\tt FeynHiggs} would have yielded 
$\Mh \approx 123.4 \pm 2.7 \gev$.
This point is also compatible
with the 68\%~CL limit from \bmm. 
The 95\%~CL
upper limit on \bmm\ requires $m_{1/2} \gsim 700 \gev$,
already placing a SUSY interpretation of \gmt\ ``beyond reach'',
and the ATLAS 20/fb \ETslash\ search requires $m_{1/2} > 840 \gev$.

In the upper left corner of the plane, we again see a stop LSP region
with a stop-coannihilation strip of acceptable relic density due
running along its side. As in the case of the $\tb = 10$, the strip as shown here
corresponds to values of \Mh\ that are too low. However, at larger $m_0$, this too would
be acceptable. At $m_{1/2} = 1500 \gev$ and $m_0 =  3750 \gev$, for example, we
find $\Mh \simeq 124 \pm 2 \gev$ with {\tt FeynHiggs~2.10.0},
whereas {\tt FeynHiggs~2.8.6} would have yielded $\Mh \la 120 \gev$ albeit with an uncertainty
of $\pm 5 \gev$.
Thus, in the CMSSM with $\tb = 30$ and $A_0 = 2.5 m_0$
there are two regions of compatibility with the LHC measurement of \Mh\
once the improved {\tt FeynHiggs~2.10.0} calculation of \Mh\ is taken into account.

\begin{figure}[hbt]
\begin{center}
\begin{tabular}{c c}
\includegraphics[height=8cm]{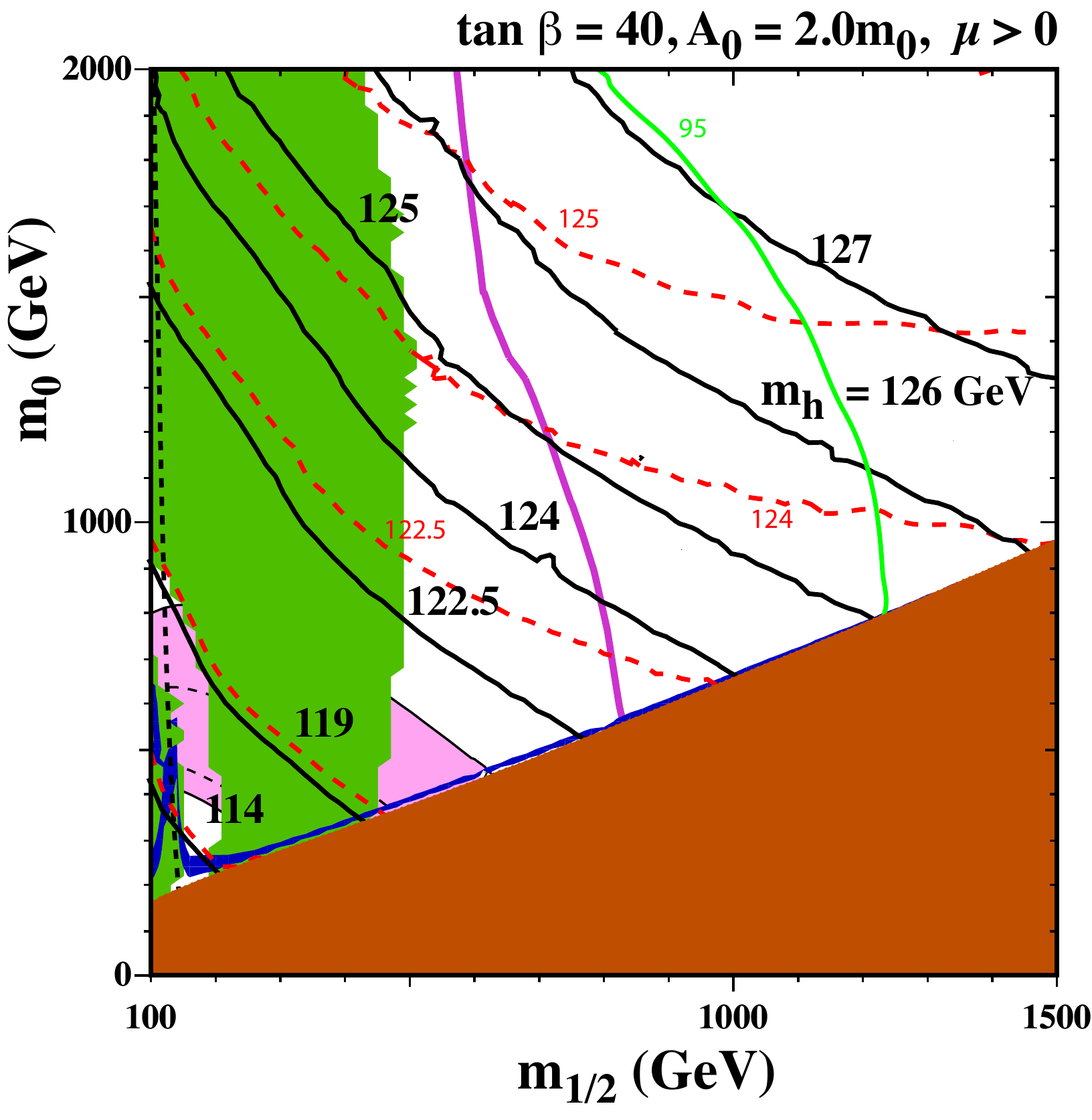} &
\includegraphics[height=8cm]{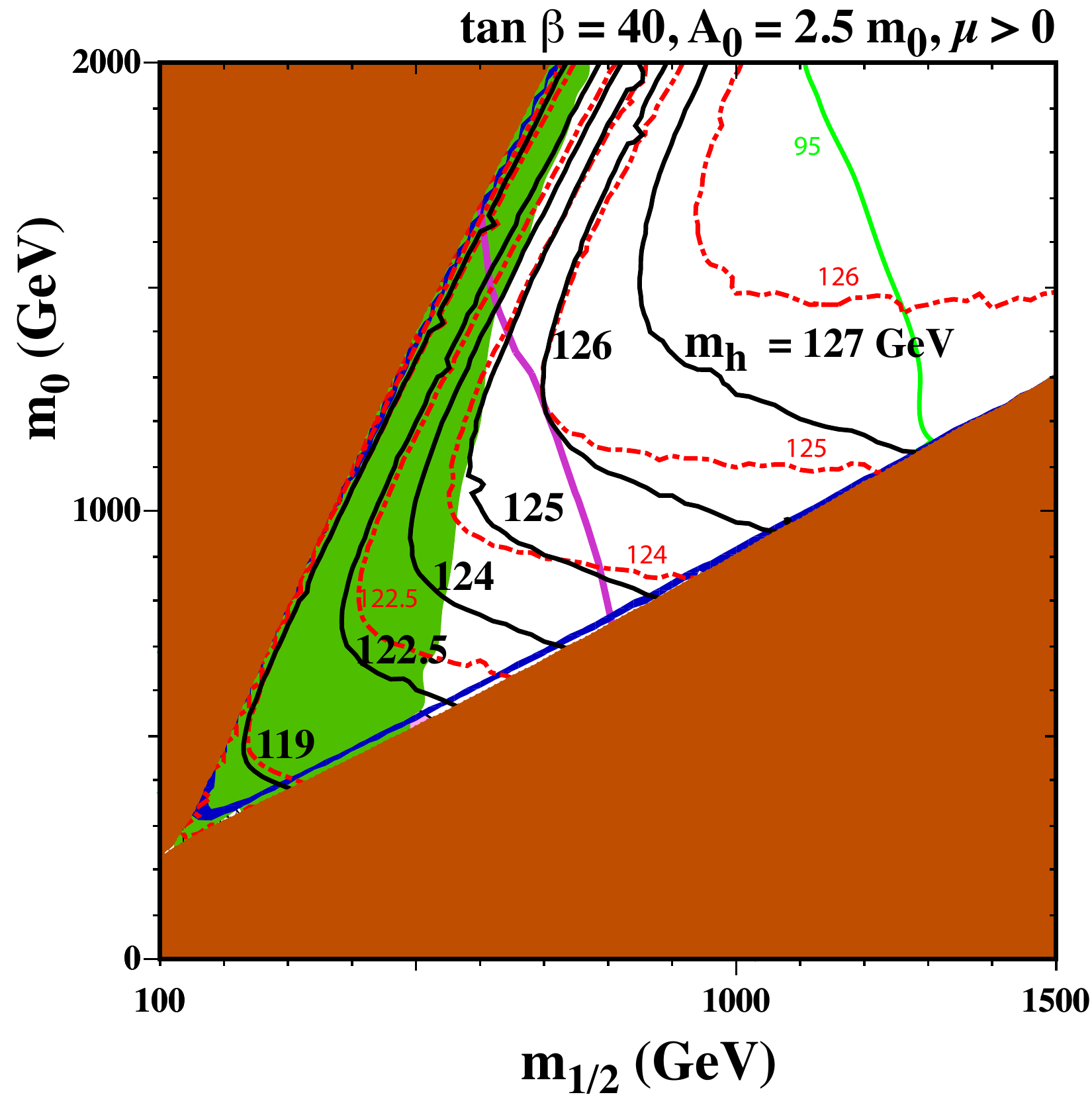} \\
\end{tabular}
\end{center}   
\caption{\label{fig:40}\it
The allowed regions in the $(m_{1/2}, m_0)$ planes for $\tb = 40$ and $A_0 = 2 m_0$
(left), $\tb = 40$ and $A_0 = 2.5 m_0$ (right). The line styles and shadings are
described in the text. When $\tb = 40$, consistency is found only if the improved {\tt FeynHiggs~2.10.0}
code is used, for the $A_0 = 2 m_0$ case.
}
\end{figure}

Fig.~\ref{fig:40} displays some analogous $(m_{1/2}, m_0)$ planes for $\tb = 40$. 
For $A_0 = 0$ (not shown), the plane would be qualitatively similar to that with $\tb = 30$, though
the constraint from \bmm\ would be much stronger. In this case,
the 95\%~CL constraint would intersect the coannihilation strip at roughly $m_{1/2} = 950 \gev$.
Instead, we show results for both $A_0 = 2 m_0$ and $2.5 m_0$.
In the case $A_0 = 2 m_0$
(left), we see that the \bmm\ 95\%~CL constraint allows
only a small section of the stau-coannihilation strip with $m_{1/2} \sim 1200 \gev$.
(The 68\% limit is at significantly higher
values of $m_{1/2}$, well past the endpoint of the coannihilation strip.)
In this case, the \bmm\ constraint is significantly stronger
than the LHC \ETslash\ constraint, and much of the region with $m_{1/2} < 500 \gev$
is also excluded by $b \to s \gamma$. 
Whereas the previous version of {\tt FeynHiggs} would have
yielded $\Mh < 123.3 \pm 2.6 \gev$ near the tip of the
stau-coannihilation strip, the improved {\tt FeynHiggs~2.10.0} calculation yields $\Mh \sim 125.0 \pm 1.1 \gev$
in this region, so it may now also be considered compatible with all
the constraints (except \gmt).

In the right panel of Fig.~\ref{fig:40}, we show the case of $\tb = 40$ and $A_0 = 2.5 m_0$.
In this case, the \bmm\ constraint also is only compatible with
the endpoint of the stau-coannihilation strip, which is now at $m_{1/2} \sim 1250 \gev$,
where the Higgs mass computed with {\tt FeynHiggs~2.10.0} is as large as 127 GeV.
(Once again, the LHC \ETslash\ constraint on $m_{1/2}$ is weaker, as is the $b \to s \gamma$ constraint.)
In the upper left corner at $m_0 \gg m_{1/2}$, we again see a stop LSP region, and a
stop-coannihilation strip running along its side. The part of the strip shown is excluded 
by $b \to s \gamma$, but compatibility is found at larger $m_0$.
For $m_{1/2} = 1500 \gev$ and $m_0 = 4050 \gev$, the stop-coannihilation strip
is compatible with both constraints on $B$ decays, but {\tt FeynHiggs~2.10.0} yields
$\Mh = 120 \gev$, albeit with a larger uncertainty $\sim 2 \gev$.

We have also considered the larger value $\tb = 55$, but find in this case
that the \bmm\ constraint is incompatible with the dark matter constraint.


\subsection{The NUHM1}

In the NUHM1, universality of the input soft SUSY-breaking gaugino, squark and slepton
masses is retained, and the corresponding contributions to the Higgs multiplets are allowed to be
different but assumed to be equal to each other. In this case, there is an additional free parameter compared
with the CMSSM, which allows one to choose either the Higgs superpotential mixing parameter $\mu$
or the pseudoscalar mass $\MA$ as a free parameter while satisfying the
electroweak vacuum conditions. 
Here and in the following we neglect the \gmt\ constraint, which is
compatible with the ATLAS~\ETslash\ searches only at around the $\pm 2.5$-$3\,\sigma$
level. 

\begin{figure}
\vspace{-2cm}
\begin{center}
\begin{tabular}{c c}
\includegraphics[height=8cm]{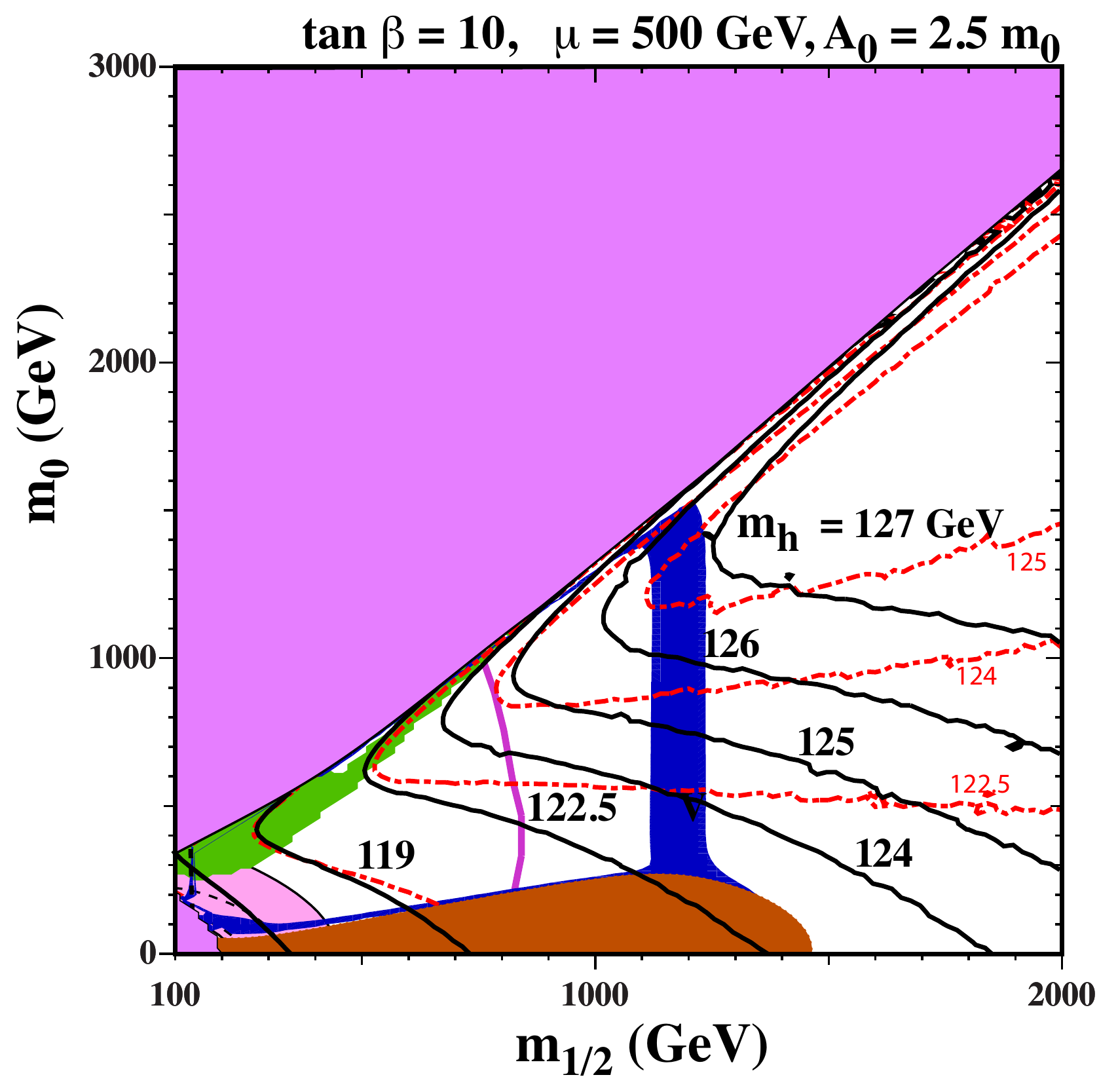} &
\includegraphics[height=8cm]{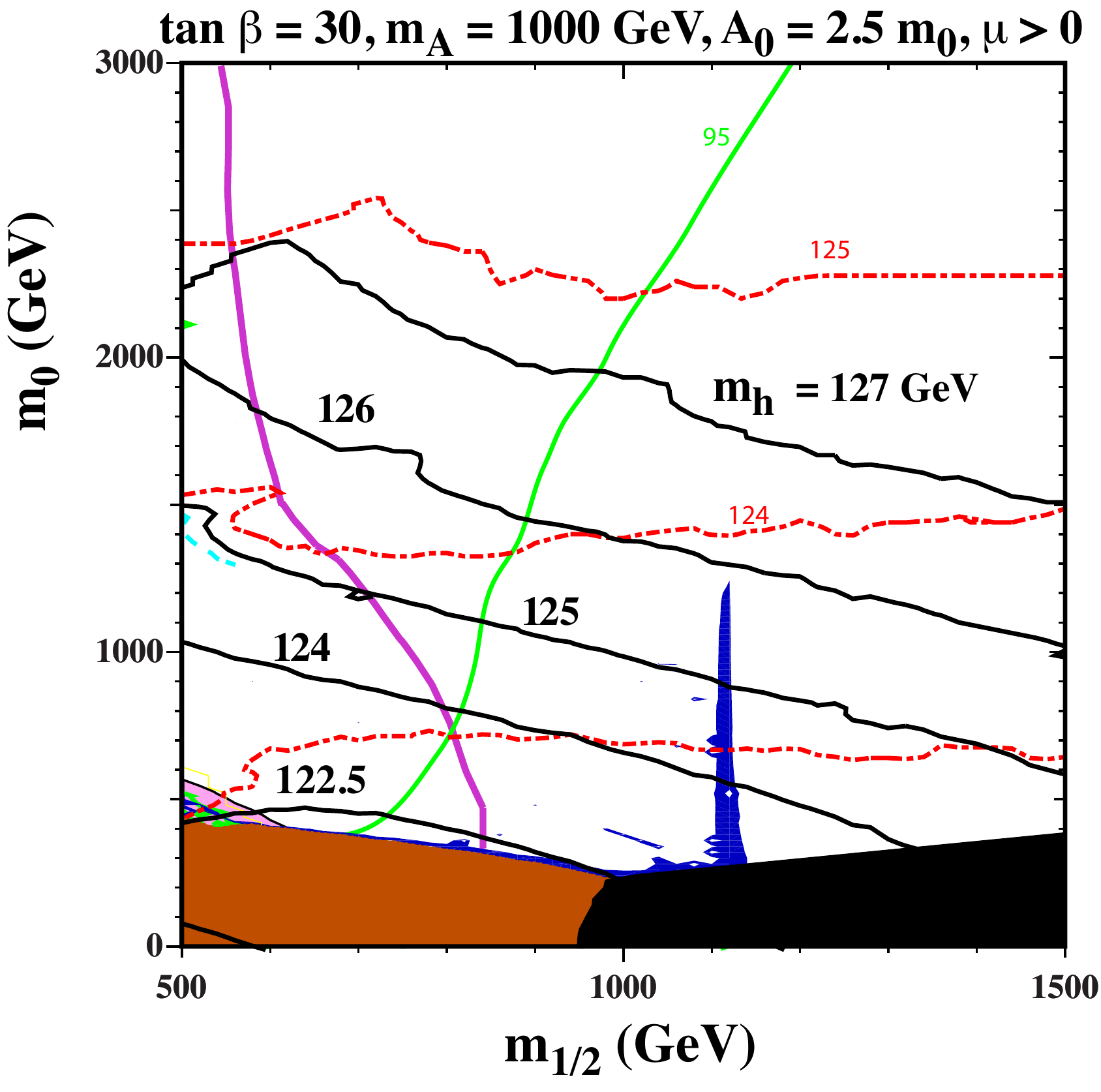} \\
\end{tabular}
\end{center}   
\begin{center}
\begin{tabular}{c c}
\includegraphics[height=8cm]{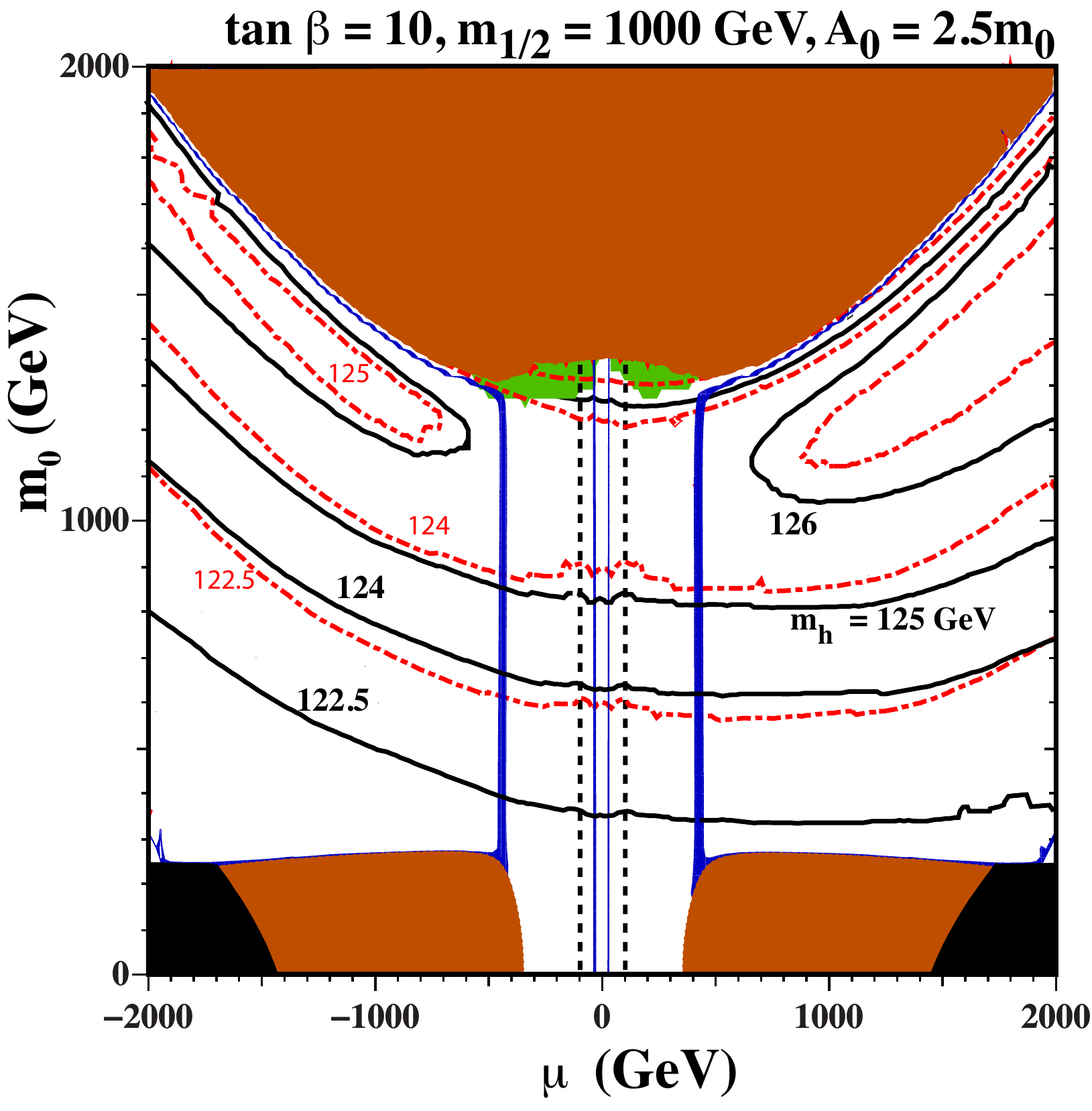} &
\includegraphics[height=8cm]{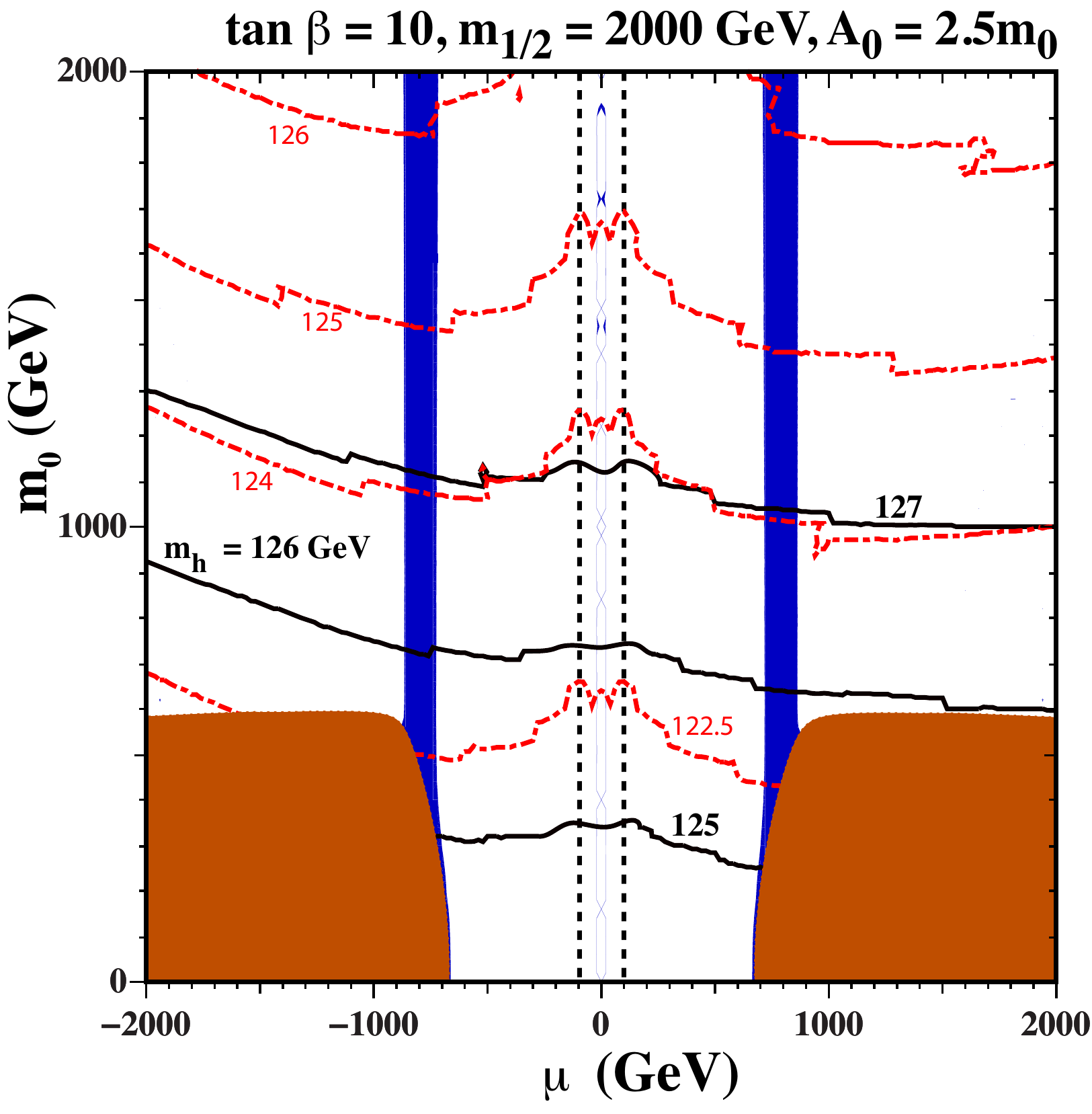} \\
\end{tabular}
\end{center}   
\caption{\label{fig:NUHM}\it
Examples of parameter planes in the NUHM1.
Two $(m_{1/2}, m_0)$ planes shown in the upper panels have $A_0 = 2.5 m_0$ for 
$\tan \beta = 10$ and $\mu = 500 \gev$ (left) and $\tan \beta = 30$ and $\MA = 1000 \gev$ (right).
Also shown are  $(\mu, m_0)$ planes with $\tb = 10$ and $m_{1/2} = 1000 \gev$
(lower left) and $m_{1/2} = 2000 \gev$ (lower right). In all the panels there are regions of consistency
with all the experimental constraints if the improved {\tt FeynHiggs~2.10.0} code is used.
}
\end{figure}

The upper left panel of Fig.~\ref{fig:NUHM} displays the NUHM1
$(m_{1/2}, m_0)$ plane for $\tb = 10, A_0 = 2.5 m_0$ and $\mu = 500 \gev$. In this case, we
see that the stau-coannihilation strip at low $m_0$ is connected to the focus-point strip by a broader
(dark blue) band with $m_{1/2} \sim 1200 \gev$ that is compatible with the astrophysical dark matter constraint. In this band, the
composition of the LSP has a substantial Higgsino admixture that brings the relic density down
into the astrophysical range, and its location depends on the assumed value of $\mu$.
The value chosen here, $\mu = 500 \gev$, places this band beyond the ATLAS 20/fb \ETslash\
limit, and the \bmm\ constraint is not important for this value of $\tb$.
Furthermore, we see from the \Mh\ contours that all this band is compatible with the Higgs
mass measurement if the improved code {\tt FeynHiggs~2.10.0} is used.
Only the upper part of this strip would have appeared consistent if the previous version of
{\tt FeynHiggs} had been used. This example shows that the freedom to vary $\mu$ within the NUHM1
opens up many possibilities to satisfy the experimental constraints, e.g., a lower value of $\tb$
than was possible in the CMSSM.

The upper right panel of Fig.~\ref{fig:NUHM} displays the
$(m_{1/2}, m_0)$ plane for $\tb = 30, A_0 = 2.5 m_0$ and fixed $\MA = 1000 \gev$~\footnote{Here
and in the lower left panel, in
the black shaded region the LSP is a charged slepton other than the lighter stau.}. In this case there is
a spike at $m_{1/2} \sim 1100 \gev$ in which the dark matter density is brought down into the range
allowed by astrophysics and cosmology by rapid LSP annihilations into the heavy Higgs bosons $H/A$,
a mechanism that operates whenever $\mneu1 \sim \MA/2$, namely $\sim 500 \gev$ in this case. All
of the spike is comfortably consistent with the ATLAS 20/fb \ETslash\ constraint and the upper limit
on \bmm. We
see that in the upper part of this spike {\tt FeynHiggs~2.10.0} yields a nominal value of $\Mh \in (125, 126) \gev$,
an increase of about 1.5 GeV over {\tt FeynHiggs~2.8.6},
but lower parts of the spike may also be consistent with the LHC Higgs mass measurement, given the
theoretical uncertainties. On the other hand, only limited
consistency in the lower part of the strip would have been found with the 
previous version of {\tt FeynHiggs}. This example shows that the freedom to vary $\MA$ within the NUHM1
opens up many possibilities to satisfy the experimental constraints.

In the lower left panel of Fig.~\ref{fig:NUHM} we display a different type of slice
through the NUHM1 parameter space, namely a $(\mu, m_0)$ plane for fixed
$\tb = 10, m_{1/2} = 1000 \gev$ and $A_0 = 2.5 m_0$. With this choice of
$m_{1/2}$, the ATLAS 20/fb \ETslash\ constraint is automatically satisfied thoughout
the plane, and with this choice of $\tb$ the \bmm\
constraint is also satisfied everywhere. We see two near-vertical dark blue bands
where the relic LSP density falls within the cosmological range, again because
of a large admixture of Higgsino in the LSP composition associated with the
the near-degeneracy of two neutralino mass eigenstates. These bands stretch
between a stop LSP region at large $m_0$ and a stau LSP region at low $m_0$,
which is flanked by charged slepton LSP regions at large $|\mu|$. We see that over
much of this plane the value of \Mh\ calculated with {\tt FeynHiggs~2.10.0}
is $\sim 1 \gev$ higher than the {\tt 2.8.6} value. The upper parts of the dark blue 
bands again yield a nominal value of $\Mh \in (125, 126) \gev$, and much of the
rest of the bands may be compatible within the theoretical uncertainties. 

The same is true in the lower right panel of Fig.~\ref{fig:NUHM}, where we display
an analogous $(\mu, m_0)$ plane for $\tb = 10, m_{1/2} = 2000 \gev$ and $A_0 = 2.5 m_0$.
Here we see that the stau LSP regions have expanded to larger $m_0$, and there are again
near-vertical dark matter bands rising from them, whilst the stop LSP region has receded to larger $m_0$.
In general, values of \Mh\ are larger than previously, with {\tt FeynHiggs~2.10.0} yielding
nominal values $\ga 127 \gev$ for $m_0 > 1000 \gev$. This is roughly 3 GeV higher than found in
{\tt FeynHiggs~2.8.6}. In this case, values of \Mh\
as low as 125~GeV are attained only at the lower tips of the dark matter bands,
very close to the stau LSP region with $m_0 \sim 300 \gev$. However, the entire
bands are probably compatible with the LHC measurement of \Mh\ when the theoretical
uncertainties are taken into account.

We conclude from the analysis in this Section that values of $\Mh \sim
125$ to 126~GeV are unexceptional in the NUHM1, and possible, e.g., for
smaller values of $\tb$ than in the CMSSM, though disfavouring a
supersymmetric interpretation of \gmt.


\subsection{The NUHM2}

\begin{figure}
\begin{center}
\begin{tabular}{c c}
\includegraphics[height=8cm]{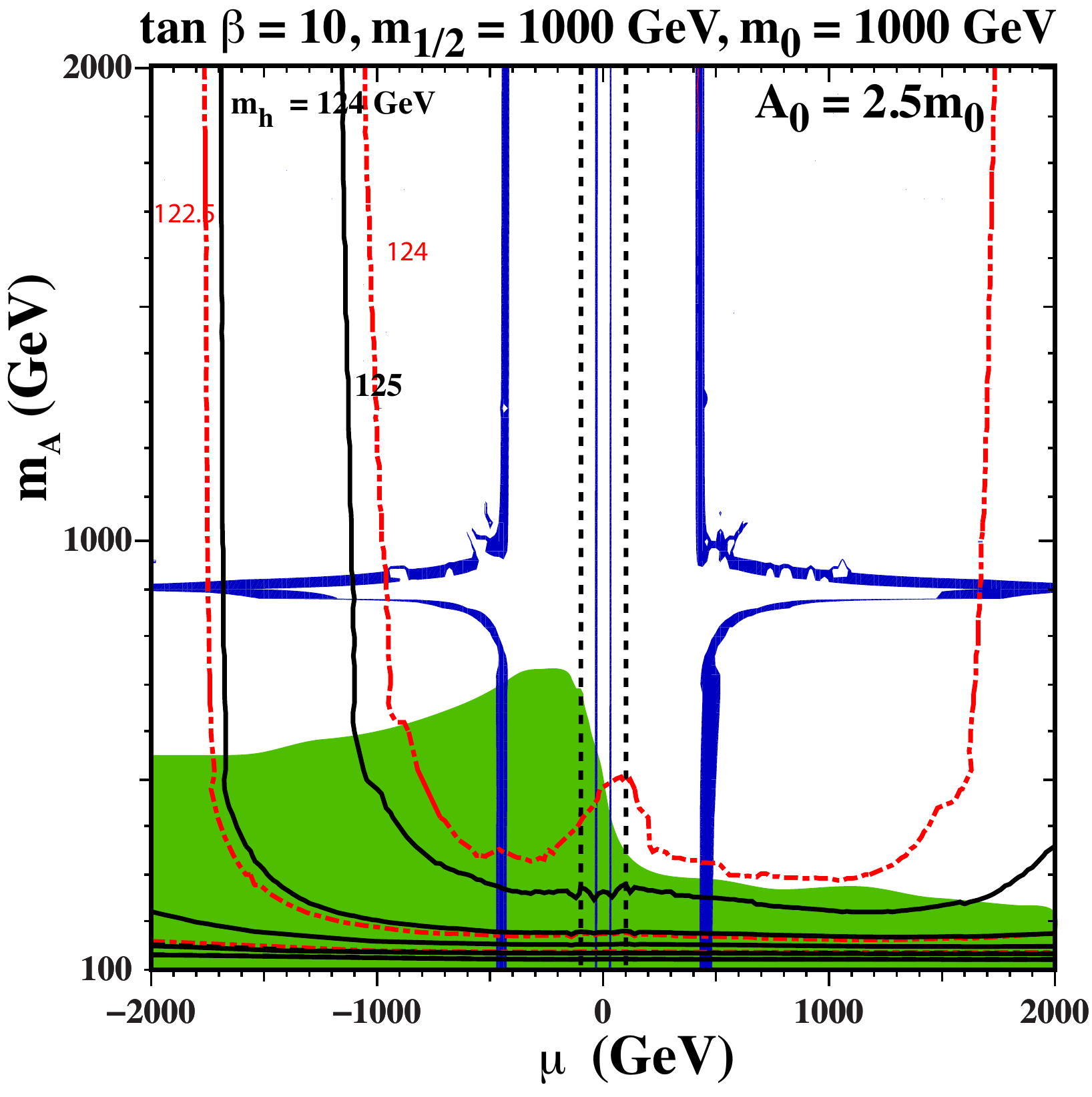} &
\includegraphics[height=8cm]{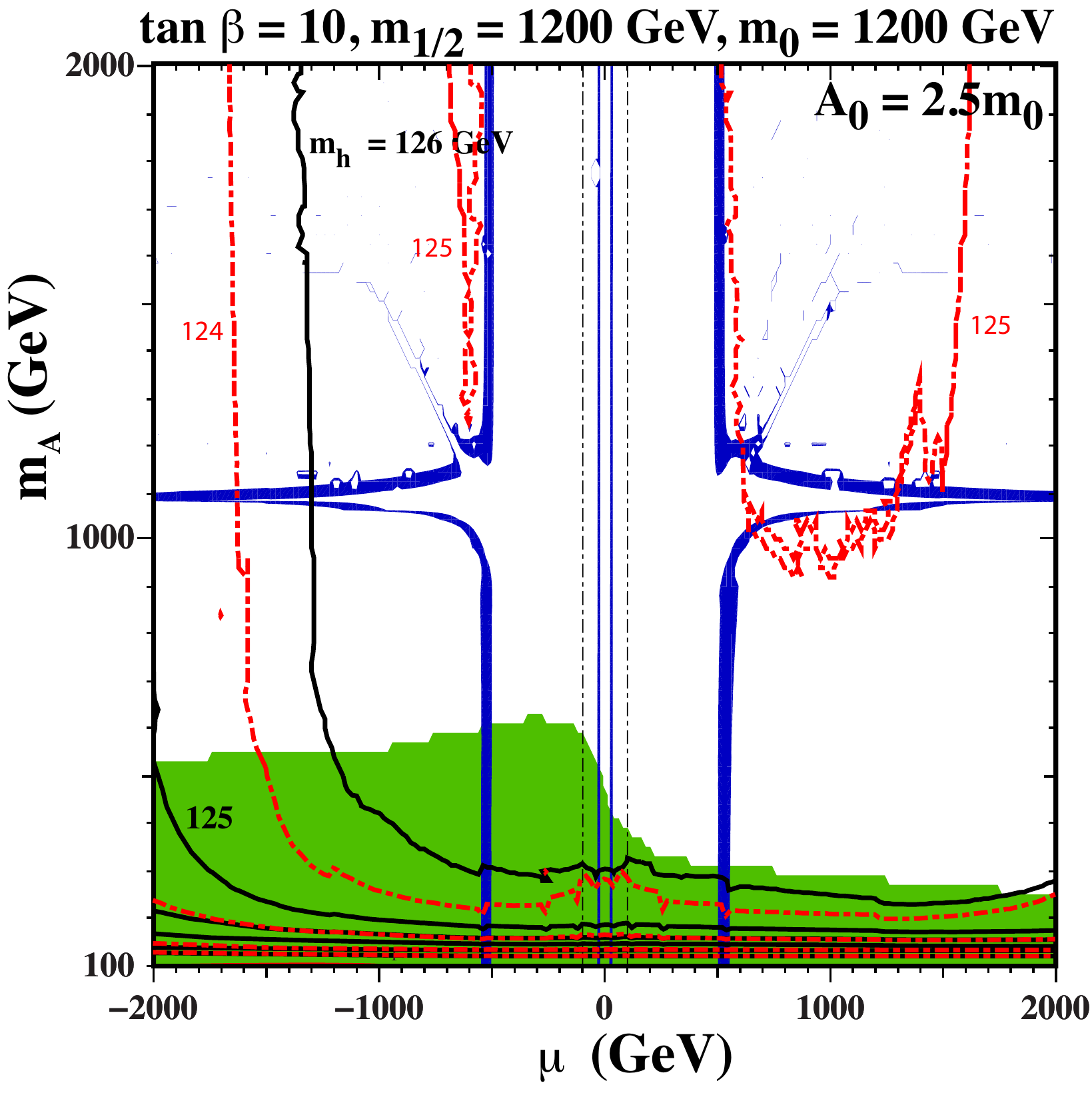} \\
\end{tabular}
\end{center}   
\caption{\label{fig:NUHM2}\it
Examples of $(\mu, \MA)$ plane in the NUHM2 for $tan beta = 10$
and $A_0 = 2.5 m_0$, with $m_{1/2} = m_0 = 1000 \gev$
(left), and with $m_{1/2} = m_0 = 2000 \gev$ (right). Using the improved {\tt FeynHiggs~2.10.0} code,
consistency with the measured value of \Mh\ is found over all the
dark matter bands in both panels.
}
\end{figure}

In the NUHM2, the soft SUSY-breaking contributions to the masses
of the two Higgs multiplets are allowed to vary independently, so there are
two additional parameters compared to the CMSSM, which may be taken as
$\mu$ and $\MA$. Fig.~\ref{fig:NUHM2} displays
illustrative $(\mu, \MA)$ planes for fixed values of the other parameters
$\tb = 10, A_0 = 2.5 m_0$ and $m_{1/2} = m_0 = 1000 \gev$ (left), 
$m_{1/2} = m_0 = 1200 \gev$ (right).
We see immediately that the $b \to s \gamma$ constraint is stronger for $\mu <0$
(which is one of the reasons that more studies have been made of models with $\mu >0$)
and that \Mh\ is generally larger for $\mu > 0$ than for $\mu < 0$, if equal values of
the other model parameters are chosen. The vertical dark matter strips
correspond to large Higgsino admixtures, as in the NUHM1 examples discussed earlier,
and the horizontal funnels are due to enhancement of LSP annihilation by direct-channel
$H/A$ poles: these move to higher (lower) $\MA$ for larger (smaller) $m_{1/2}$,
as seen by comparing the left and right panels of Fig.~\ref{fig:NUHM2}.

All the dark matter-compatible points in the left panel would correspond
to values of \Mh\ consistent with the experimental measurements within the
theoretical uncertainties. In this case,
the shift in \Mh\ from {\tt FeynHiggs~2.8.6} to {\tt FeynHiggs~2.10.0} is about 1 GeV 
at $m_{1/2} = m_0 = 1000 \gev$ and somewhat larger at higher $m_{1/2}, m_0$
as seen in the right panel.
In the right panel we see that typical nominal  {\tt FeynHiggs~2.10.0}
values of \Mh\ are larger than the measured value, though
they are consistent with experiment, given the theoretical uncertainties.

\subsection{mSUGRA}

Finally, we consider a scenario that is more restrictive than the CMSSM, namely
minimal supergravity (mSUGRA). In this case, there is a universal input scalar
mass $m_0$ equal to the gravitino mass $m_{3/2}$, and the soft bilinear and
trilinear soft SUSY-breaking masses are related by $A_0 = (B_0 + 1) m_0$,
see \cite{AbdusSalam:2011fc} for a review.
The first constraint means that we do not have the luxury of assuming $m_{3/2}$
to be arbitrarily large, and there are regions of the $(m_{1/2}, m_0)$ plane
where the LSP is necessarily the gravitino. The relation between $A_0$ and $B_0$
implies that $\tb$ is determined at any point in the $(m_{1/2}, m_0)$ plane
once $A_0$ is fixed.

\begin{figure}[h]
\begin{center}
\includegraphics[height=8cm]{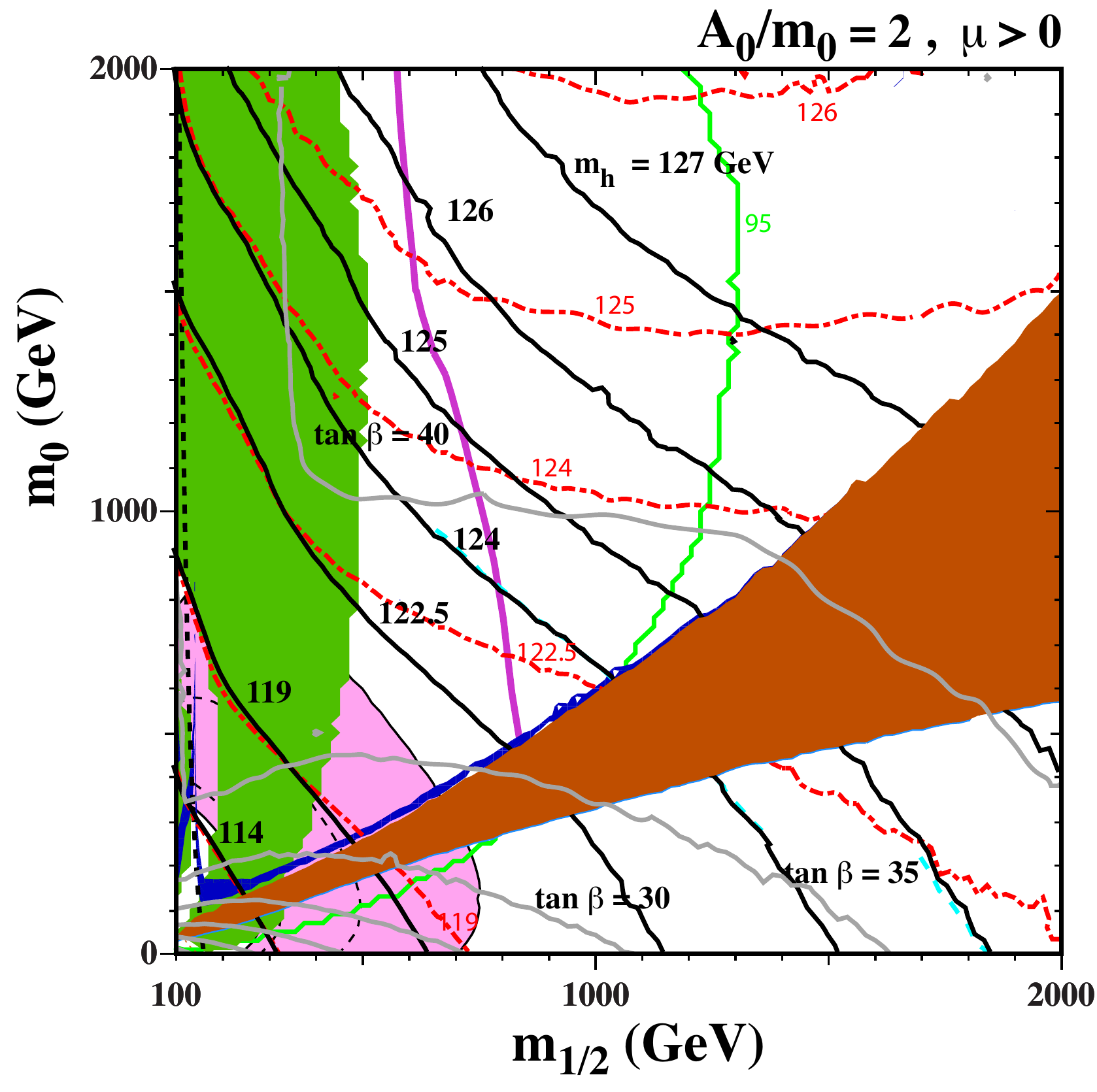} 
\end{center}   
\caption{\label{fig:mSUGRA}\it
The allowed regions in the $(m_{1/2}, m_0)$ plane in a mSUGRA model with
$A_0/m_0 = 2$. In addition to the line and shade descriptions found in the text,
shown here are labeled solid grey contours showing the derived value of
$\tb$.
Using the improved {\tt FeynHiggs~2.10.0} code,
consistency with the measured value of \Mh\ is found near the tip of the
stau-coannihilation strip.
}
\end{figure}

Both these features are visible in Fig.~\ref{fig:mSUGRA}, where the $(m_{1/2}, m_0 = m_{3/2})$
plane for $A_0 = 2 m_0$ and $\mu > 0$ exhibits (grey) contours of $\tb$ and a wedge
where the LSP is the lighter stau, flanked by a neutralino LSP region at larger $m_0 = m_{3/2}$
and a gravitino LSP region at smaller $m_0 = m_{3/2}$. The ATLAS 20/fb \ETslash\ search is
directly applicable only in the neutralino LSP region, and requires reconsideration in the
gravitino LSP region. In addition, in this region there are important astrophysical
and cosmological limits on long-lived charged particles (in this case staus), that we do not
consider here, so we concentrate on the neutralino LSP region above the stau LSP wedge.
The ATLAS 20/fb \ETslash\ constraint intersects the dark matter coannihilation strip just above
this wedge where $m_{1/2} \sim 850 \gev$, and the \bmm\ constraint
intersects the coannihilation strip at $m_{1/2} \sim 1050 \gev$, whereas the tip of the
strip is at $m_{1/2} \sim 1250 \gev$. 
In this section of the coannihilation strip the nominal
value of \Mh\ provided by the improved {\tt FeynHiggs~2.10.0} calculation is $\in (124, 125) \gev$, compatible
with the experimental measurement within the theoretical uncertainties due to the 1-2 GeV shift
in \Mh\ found in this new version of {\tt FeynHiggs}.

\section{Higgs Mass Predictions from Global Fits within the CMSSM and NUHM1}
\label{sec:mastercode}

We saw in previous Sections that different calculations of \Mh\ may differ
significantly, particularly at large values of $m_{1/2}$ and/or $m_0$. 
With the improved \Mh\ calculation in {\tt FeynHiggs} {\tt 2.10.0}, the
theory uncertainty has now been reduced to allow more precise
\Mh\ evaluations also for larger values of the relevant SUSY
parameters. Taking this into account, we
found regions in the CMSSM that were compatible with the LHC measurement
of \Mh\ and other constraints when the improved {\tt FeynHiggs~2.10.0}
code is used, as well as broader possibilities for compatibility
in the NUHM1 and NUHM2. In this
Section we consider the possible implications for global fits to SUSY model
parameters that include \Mh\ in the construction of the global likelihood
function, concentrating for definiteness on the CMSSM and NUHM1 fits
presented in~\cite{mc8}.

\begin{figure}
\centerline{
\mbox{}\hspace{10mm}
\includegraphics[height=7.2cm]{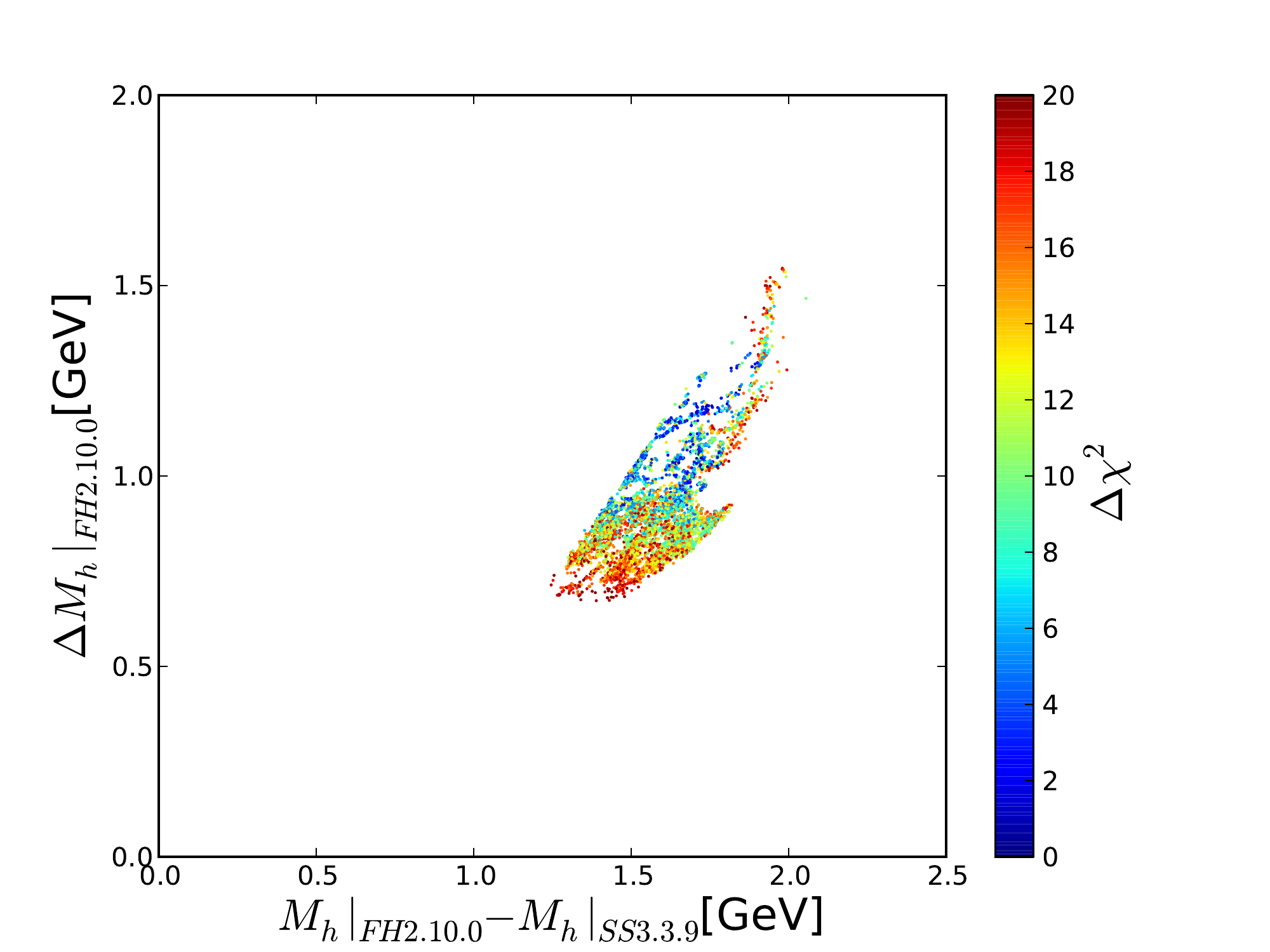}
\mbox{}\hspace{-12mm}
\includegraphics[height=7.2cm]{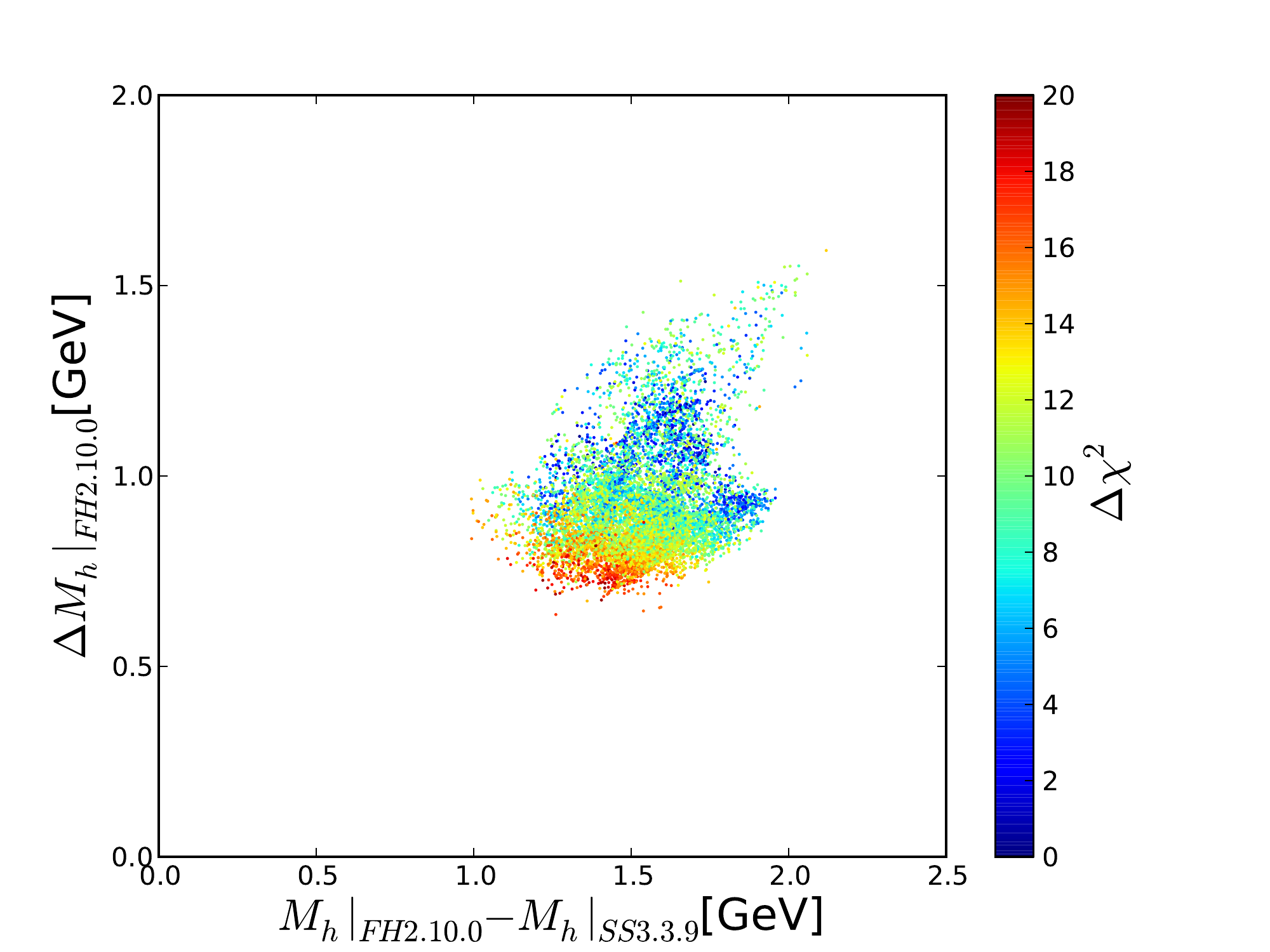} 
}
\caption{\label{fig:fhssfits}\it
Scatter plots of 10000 points each selected randomly from
scans~\protect\cite{mc8} in the CMSSM (left) and the NUHM1 (right), 
displayed in 
$(\Mh|_{{\rm FH}2.10.0} - \Mh|_{{\rm SS}3.3.9}, \Delta \Mh|_{{\rm FH}2.10.0})$ 
planes and colour-coded according to their $\chi^2$ values.
}
\end{figure}

In the following we will compare {\tt FeynHiggs~2.10.0} with 
{\tt SoftSusy~3.3.9}. While the higher-order corrections included in 
{\tt FeynHiggs~2.10.0} are more complete than those in {\tt SoftSusy}, a very
large discrepancy between the 
two codes would indicate a parameter region that is potentially unstable
under higher-order corrections in at least one of the codes.
Fig.~\ref{fig:fhssfits} displays planes of 
$\Mh|_{{\rm FH}2.10.0} - \Mh|_{{\rm SS}3.3.9}$ 
vs.\ the theoretical uncertainty $\Delta \Mh|_{{\rm FH}2.10.0}$ estimated
within {\tt FeynHiggs~2.10.0} (see~\cite{newFH} for details),
displaying 10000 points chosen randomly from the samples in~\cite{mc8}
(but with an upper limit on $\Delta \chi^2 < 20$ to concentrate on the
parts of parameter space of most phenomenological relevance) for the CMSSM (left panel)
and the NUHM1 (right panel). The points are colour-coded
according to the differences found in~\cite{mc8} between their $\chi^2$
values and those of the best-fit points in the CMSSM and NUHM1, respectively,
with low-$\Delta \chi^2$ points in blue and high-$\Delta \chi^2$ points in red.

The differences between the two codes are found in the region of
$|\Mh|_{{\rm FH}2.10.0} - \Mh|_{{\rm SS}3.3.9}| = 1.0 - 2.0 \gev$ with a
theoretical uncertainty prediction (for only the
\fh\ calculation) between $\sim 0.6$ and $\sim 1.5$. The consistent
difference between the two codes can be attributed to the more complete
inclusion of higher-order corrections in \fh, which is reflected in the
fact that the difference often exceeds the \fh\ theory uncertainty. 
On the other hand, no phenomenologically relevant parameter points are
found with an unexpectedly large difference between the two codes. This
indicates that the relevant parameter regions are not located in parts of
the CMSSM/NUHM1 parameter space that lead to an unstable
\Mh\ evaluation. This supports the viability of the constraints imposed by \Mh\ on these
models.

\begin{figure}
\begin{center}
\includegraphics[height=7cm]{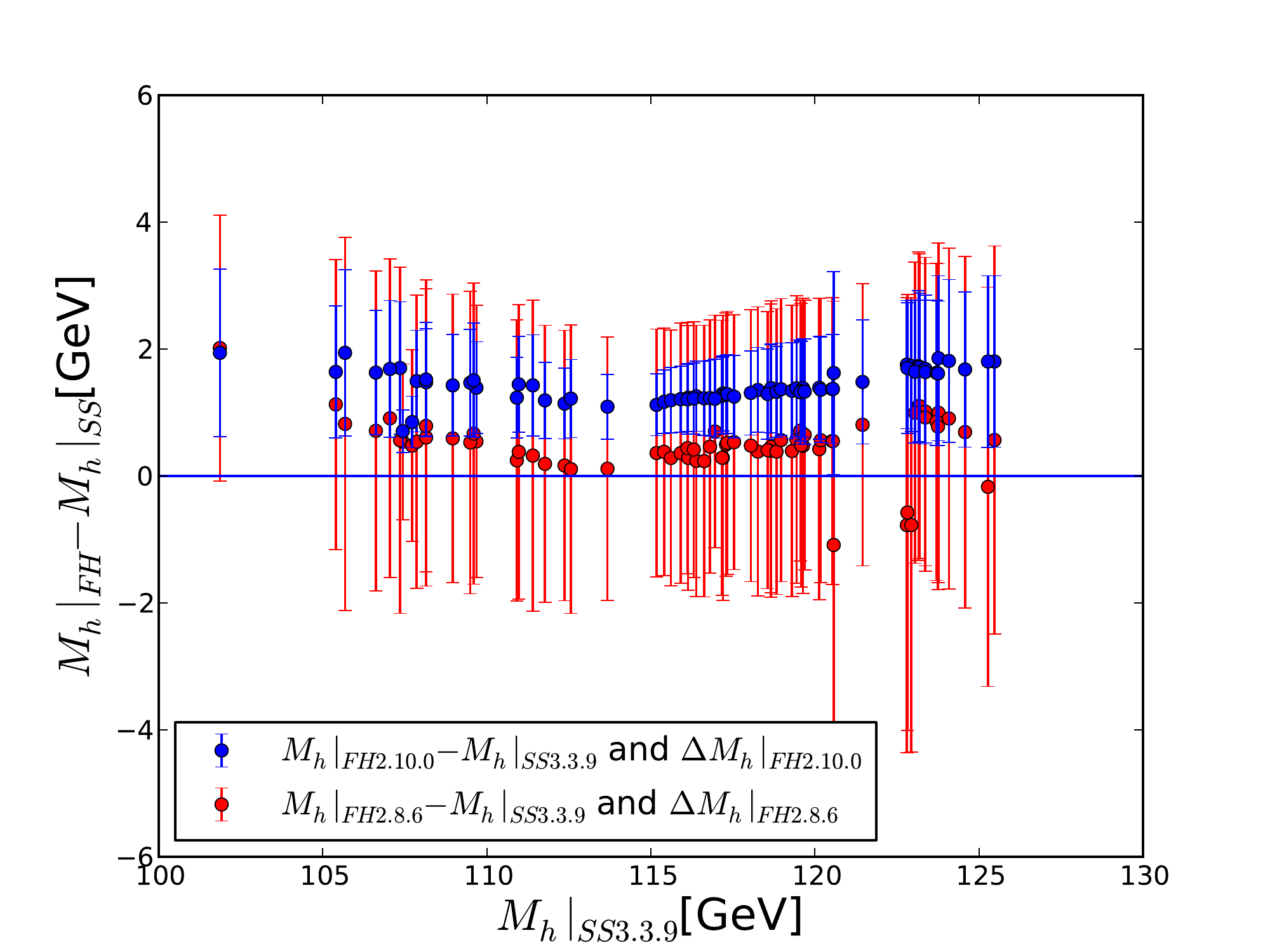}
\end{center}   
\caption{\label{fig:differences}\it
Values of $\Mh|_{{\rm FH}2.10.0} - \Mh|_{{\rm SS}3.3.9}$ (in dark blue) 
and of $\Mh|_{{\rm FH}2.8.6} - \Mh|_{{\rm SS}3.3.9}$ (in red) plotted against
$\Mh|_{{\rm SS}3.3.9}$,
for 100 CMSSM points from the sample
in~\cite{mc8} that have the lowest $\chi^2$ for each bin in \Mh.
The vertical lines indicate the respective \Mh\ uncertainty
calculations as evaluated by \fh.
}
\end{figure}

A similar inference can be drawn from Fig.~\ref{fig:differences}.
For this plot we have selected 100 CMSSM points from the sample
in~\cite{mc8} that have the lowest $\chi^2$ for each bin in 
$\Mh|_{{\rm SS}3.3.9}$. 
We show their values of $\Mh|_{{\rm FH}2.10.0} - \Mh|_{{\rm SS}3.3.9}$
(in dark blue) and of $\Mh|_{{\rm FH}2.8.6} - \Mh|_{{\rm SS}3.3.9}$ (in
red) on the vertical axis, 
using $\Mh|_{{\rm SS}3.3.9}$ as the horizontal axis. 
In both cases the respective \Mh\ uncerainty calculations of
\fh\ are indicated via vertical lines.
We see that both
{\tt FeynHiggs~2.10.0} and {\tt 2.8.6} yield values of \Mh\ that are
systematically larger than {\tt SoftSusy~3.3.9}. In most cases, 
$1 \gev \lsim \Mh|_{{\rm FH}2.10.0} - \Mh|_{{\rm SS}3.3.9} \lsim 2 \gev$
and $0 \lsim \Mh|_{{\rm FH}2.8.6} - \Mh|_{{\rm SS}3.3.9} \lsim 1 \gev$,
and $\Mh|_{{\rm FH}2.10.0} - \Mh|_{{\rm FH}2.8.6} \sim 1 \gev$. 
The change from version {\tt 2.8.6} to version {\tt 2.10.0}
reflects the size of the newly-included resummed corrections to $\Mh$
for a relevant part of the parameter space.

The theoretical \Mh\ uncertainty evaluated in \fh\,{\tt 2.8.6}
embraced the {\tt SoftSusy} predictions as well as the updated 
\fh\,{\tt 2.10.0} prediction for \Mh. The latter, in particular, gives
confidence that the uncertainty calculation indeed captures the missing
higher-order corrections. The new theoretical uncertainty as evaluated
using \fh\,{\tt 2.10.0} does not include, in general, the older
\fh\ prediction, nor does it include (in all cases) the {\tt SoftSusy}
prediction. This again demonstrates the effects and the relevance of the
newly-included resummed logarithmic corrections in \fh.

\section{Summary and Conclusions}

As we have shown in this paper, the improved Higgs mass calculations
provided in the improved {\tt FeynHiggs~2.10.0} code have significant implications
for the allowed parameter spaces of supersymmetric models. We have
illustrated this point with examples in the pMSSM, CMSSM, NUHM1 and NUHM2
frameworks. 

In a random scan of the pMSSM10 parameter space we exhibited the change in the Higgs mass 
$\Delta \Mh$ in {\tt FeynHiggs~2.10.0} compared to the previous version {\tt FeynHiggs~2.8.6}. 
This averages below 2~GeV for third family squark masses below 2~TeV, 
but can increase up to $\Delta \Mh\sim 5 \gev$ for $\msqt = 5 \tev$.
The update to {\tt FeynHiggs~2.10.0} is therefore particularly relevant in light
of the measured value of $\Mh$ and the strengthened LHC lower limits on sparticle masses.

The CMSSM is under strong pressure from the LHC searches
for jets + \ETslash ~events, which exclude small values of $m_{1/2}$,
the measurement of BR($B_s \to \mu^+ \mu^-$), which disfavours
large values of $\tb$, the measurement of $\Mh$, which favours
large values of $m_{1/2}$ and/or $\tb$ and positive values of
$A_0$, and the cosmological dark matter density constraint. We have
shown that these constraints can be reconciled for suitable intermediate
values of $\tb$ if {\tt FeynHiggs~2.10.0} is used to calculate $\Mh$
in terms of the input CMSSM parameters (with the exception of \gmt). 
The pressure on the CMSSM
would have been much greater if an earlier version of {\tt FeynHiggs}
had been used, which yielded lower values of $\Mh$ because it did not
include the leading and next-to-leading logarithms of type $\log(\mst/\mt)$ 
in all orders of perturbation theory as incorporated in {\tt FeynHiggs~2.10.0}.

The LHC constraints are satisfied more easily in the NUHM1 (and NUHM2),
with their one (or two) extra parameters that offer more options for
satisfying the cosmological dark matter density constraint at larger
values of $m_{1/2}$ than in the CMSSM. The extra degree(s) of freedom in the
NUHM1 (NUHM2) allow the Higgs mixing parameter $\mu$ or (and) $\MA$
to be adjusted so that a sizable Higgsino component is present increasing the
annihilation cross section, and/or allowing $\chi \chi^\pm$ and/or rapid
direct-channel $\neu1\neu1 \to H/A$
annihilation to bring  the cosmological dark matter density into the
allowed range. 
Reconciling all the constraints would have been possible already with the
earlier version of {\tt FeynHiggs}, but is easier to achieve
when the improved {\tt FeynHiggs~2.10.0} version is used.

In addition to the higher values of $\Mh$ yielded by {\tt FeynHiggs~2.10.0},
this code also provides a correspondingly reduced
estimate of the theoretical uncertainty in the mass
calculation. This must also be taken into account when analyzing the
consistency with other constraints within the CMSSM, NUHM1, NUHM2 or
any other models. Taken together, the improved mass calculations
and uncertainty estimates in {\tt FeynHiggs~2.10.0} make it a preferred
tool for the analysis of supersymmetric models.

\vspace{1cm}
\noindent{ {\bf Acknowledgments} } \\
\noindent
The work of J.E. was supported in part by
the London Centre for Terauniverse Studies (LCTS), using funding from
the European Research Council 
via the Advanced Investigator Grant 267352.
The work of K.A.O. was supported in part
by DOE grant DE--FG02--94ER--40823 at the University of Minnesota.
The work of S.H.\ is supported 
in part by CICYT (grant FPA 2010--22163-C02-01) and by the
Spanish MICINN's Consolider-Ingenio 2010 Program under grant MultiDark
CSD2009-00064. 
The work of G.W.\ was supported by the Collaborative
Research Center SFB676 of the DFG, ``Particles, Strings, and the early
Universe".


\end{document}